%% file: fan-rdg.tex
\documentclass[12pt,preprint]{emulateapj}

\def\arcs{$''$}

\def\hub{\ifmmode H_\circ\else H$_\circ$\fi}

\shorttitle{Star clusters in M31} \shortauthors{Fan, de Grijs, \& Zhou}

\begin{document}

\title{An updated catalog of M31 globular-like clusters: $UBVRI$ photometry, ages, and masses}

\author{Zhou Fan,\altaffilmark{1,2} Richard de Grijs,\altaffilmark{3} \& Xu Zhou\altaffilmark{1,2}}

\altaffiltext{1}{National Astronomical Observatories, Chinese Academy of
  Sciences, A20 Datun Road, Chaoyang District, Beijing 100012, China}

\altaffiltext{2}{Key Laboratory of Optical Astronomy, National
Astronomical Observatories, Chinese Academy of Sciences, Beijing 100012,
China}

\altaffiltext{3}{Kavli Institute for Astronomy and Astrophysics,
Peking University, Yi He Yuan Lu 5, Hai Dian District, Beijing 100871,
China}

\email{zfan@nao.cas.cn}

\begin{abstract}
We present an updated $UBVRI$ photometric catalog containing 970
objects in the field of M31, selected from the Revised Bologna Catalog
(RBC v.4.0), including 965, 967, 965, 953, and 827 sources in the
individual $UBVRI$ bands, respectively, of which 205, 123, 14,
126, and 109 objects do not have previously published photometry.
Photometry is performed using archival images from the Local Group
Galaxies Survey, which covers 2.2 deg$^2$ along the major axis of
M31. Detailed comparisons show that our photometry is fully consistent
with previous measurements in all filters. We focus on 445 confirmed
`globular-like' clusters and candidates, comprising typical globular
and young massive clusters. The ages and masses of these objects are
derived by comparison of their observed spectral-energy distributions
with simple stellar population synthesis. Approximately half of the
clusters are younger than 2 Gyr, suggesting that there has been
significant recent active star formation in M31, which is consistent
with previous results. We note that clusters in the halo ($r_{\rm
projected}>30$ kpc) are composed of two different components, older
clusters with ages $>10$ Gyr and younger clusters with ages around 1
Gyr. The spatial distributions show that the young clusters ($<2$ Gyr)
are spatially coincident with the galaxy's disk, including the `10 kpc
ring,' the `outer ring,' and the halo of M31, while the old clusters
($> 2$ Gyr) are spatially correlated with the bulge and halo. We also
estimate the masses of the 445 confirmed clusters and candidates in
M31 and find that our estimates agree well with previously published
values. We find that none of the young disk clusters can survive the
inevitable encounters with giant molecular clouds in the galaxy's disk
and that they will eventually disrupt on timescales of a few
Gyr. Specifically, young disk clusters with a mass of $10^4 M_\odot$
are expected to dissolve within 3.0 Gyr and will, thus, not evolve to
become globular clusters.
\end{abstract}


\section{Introduction}
\label{s:intro}

Star clusters comprise an important stellar population component of
galaxies and their age distributions trace the main events in the
formation and evolution of their host galaxies. For a long time, star
clusters were thought of as members of two distinct types,
specifically open and globular clusters (OCs, GCs). OCs are young, not
very massive, faint, diffuse, and usually located in galactic disks,
quite contrary to the nature of GCs, which are old, massive, luminous,
centrally concentrated, and usually located in the haloes of their
host galaxies. However, this simplistic picture has been changing
since the discovery of young massive star clusters (YMCs) in many
galaxies, including the Milky Way \citep{asc07a,asc07b}, M31
\citep{bar09,ma09,cald09,per09,per10,hodge}, M82 \citep{mccr}, and NGC
1140 \citep{moll}. YMC properties span those of both OCs and GCs, with
typical masses ($>10^4 M_{\odot}$) greater than those of (most) OCs
and young ages ($< 1$ Gyr), quite different from present-day GCs, so
that they are often considered candidate proto-GCs. The new category
of YMCs renders cluster classification blurred and difficult. In this
paper, we use the term `globular-like cluster' to distinguish massive
(YMCs and GCs) from less massive clusters (OCs). Since OCs are usually
faint and located in galactic disks, which makes them difficult to
study, we focus on `globular-like clusters.'

Located at a distance of $\sim$780 kpc \citep{sg98,mac01,mc05}, M31
(also known as the Andromeda galaxy) is the nearest large spiral
galaxy in our Local Group of galaxies. Therefore, it constitutes an
ideal laboratory for studies of star clusters in external
galaxies. Based on {\sl Hubble Space Telescope (HST)} Wide Field and
Planetary Camera-2 (WFPC2) images, \citet{kh07} suggested that there
may be $\sim$80,000 star clusters in the M31 disk. Most of these disk
clusters are faint OCs. The number of GCs in M31 is much
smaller. \citet{bh01} estimated their total number at $460\pm70$, while
\citet{per10} arrived at $\sim$530, with an additional $\sim$100
YMCs. To limit the scope of this paper, we will focus on the GCs and
YMCs in M31. Since GCs and YMCs are luminous, they are relatively easy
to observe and study. Studies aimed at identification, classification,
and analysis of the population of M31 globular-like clusters have been
undertaken since the pioneering work of \citet{hub32} \citep[see,
e.g.,][]{vet62,sar77,batt80,batt87,batt93,cra85,barmby}. These studies
have provided a large amount of photometric data in different
photometric systems, including photographic plates, as well as CCD,
photoelectric, and even visual photometry.

\citet{mac06} reported the discovery of eight remote GCs in the outer
halo of M31 based on deep {\sl HST}/Advanced Camera for Surveys
images. \citet{kim} found 1164 GCs and GC candidates in M31 using the
Kitt Peak National Observatory (KPNO)'s 0.9 m and the WIYN (Wisconsin,
Indiana, Yale, and the National Optical Astronomical Observatories)
3.5 m telescopes, of which 559 and 605 were previously known GCs and
newly identified GC candidates, respectively. \citet{hux08} detected
40 new GCs in the M31 halo based on Isaac Newton Telescope and
Canada-France-Hawaii Telescope data. \citet{cald09} published a new
catalog of 670 likely star clusters, stars, possible stars, and
galaxies in the field of M31, all with updated high-quality
coordinates accurate to 0.2\arcs, based on images from either the
Local Group Galaxies Survey \citep[LGGS][]{massey} or the Digitized
Sky Survey (DSS). Recently, \citet{peac10} identified M31 clusters
using images from the UK Infrared Telescope's Wide Field Camera
(WFCAM) and the Sloan Digital Sky Survey (SDSS) archives, and obtained
photometry in the SDSS $ugriz$ and the near-infrared (NIR) $K$
bands. In addition, the authors combined all identifications and
photometry of M31 star clusters from the literature with their new
sample. Their updated M31 star cluster catalog includes 416 old,
confirmed clusters, 156 young, and 373 candidate clusters. Very
recently, \citet{hodge} discovered 77 new star clusters in star-forming
regions based on {\sl HST}/WFPC2 observations. The latest and most
comprehensive M31 GC catalog -- the Revised Bologna Catalogue of M31
globular clusters and candidates \citep[RBC v.4.0, available from
http://www.bo.astro.it/M31;][]{gall04,gall06,gall07,gall09} --
contains 2045 objects in M31, including 663 confirmed star clusters,
604 cluster candidates, and 778 other objects initially thought to be
GCs but later proved to be stars, asterisms, galaxies, and H{\sc ii}
regions. The authors adopted the photometry of \citet{barmby} as their
reference and transformed other measurements to homogenize the final
photometry database. In fact, some confirmed GCs in the RBC are
probably YMCs. The updated RBC also includes the most recently
discovered star clusters and photometry from \citet{mac06},
\citet{kim}, \citet{hux08}, and \citet{cald09}.

The $\chi^2$ minimization technique used for estimating ages,
metallicities, reddening values, and masses of extragalactic star
clusters has been described in detail by, e.g., \citet{jiang03},
\citet{deg05a}, \citet{fan06}, \citet{ma07,ma09}, and
\citet{wang10}. \citet{jiang03} presented
Beijing-Arizona-Taiwan-Connecticut (BATC) photometry of 172 GCs in the
central $\sim 1$ deg$^2$ region of M31 and estimated their ages using
simple stellar population (SSP) models. \citet{deg05a} fitted the ages
of Large Magellanic Cloud star clusters based on broad-band
spectral-energy distribution (SED) fits. \citet{fan06} estimated the
ages of 91 GCs in M31 by matching BATC intermediate-band and
Two-Micron All-Sky Survey (2MASS) $JHK$ SEDs with BC03 \citep{bc03}
SSP models. \citet{ma07} determined the ages of an old M31 GC (S312)
based on GALEX near-ultraviolet, optical broad-band, 2MASS $JHK$, and
BATC photometry. Subsequently, \citet{ma09} fitted the ages of 35 GCs
in the central M31 field that were not included in \citet{jiang03}
based on BATC, 2MASS $JHK$, and GALEX data, combined with the GALEV
SSP models.  Very recently, \citet{wang10} performed photometry for
another 104 M31 GCs with BATC multicolor observations and estimated
the ages by fitting their SEDs with GALEV SSP models, revealing the
presence of young, intermediate-age, and old cluster populations in
M31.

In this paper, we first perform aperture photometry of 970 RBC objects
based on images from the LGGS. Using photometry in the $UBVRI$ bands
and $JHK$ magnitudes from the RBC, the ages and masses of the
confirmed clusters in our sample are estimated by comparing the
observed SEDs with BC03 SSP synthesis models. This paper is organized
as follows. \S \ref{s:data} describes the sample selection and $UBVRI$
photometry. In \S \ref{s:age}, we describe the SSP models used as well
as our method to estimate the cluster ages.  In \S \ref{s:mass}, we
present the clusters' mass estimates, and we summarize and conclude
the paper in \S \ref{s:sum}.

\section{Data}
\label{s:data}

\subsection{Sample}
\label{s:samp}

\begin{figure}
\centerline{
\includegraphics[scale=.45,angle=-90]{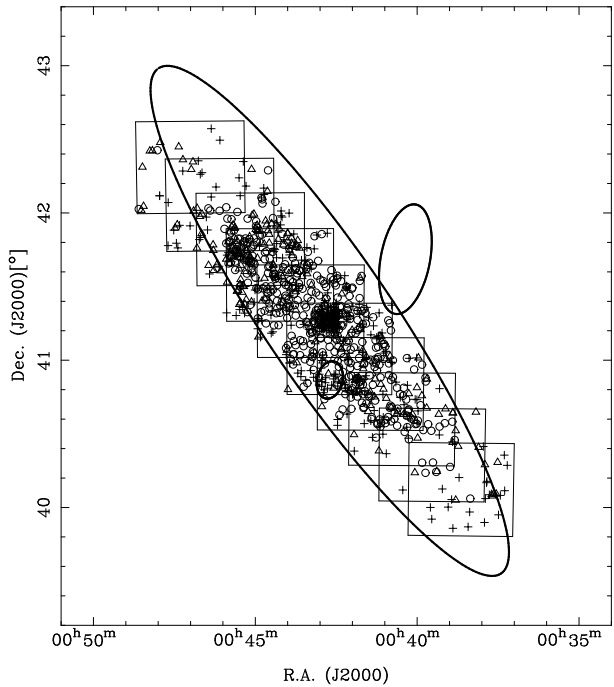}}
\caption[]{Spatial distribution of the 970 objects selected from the
  RBC v.4.0 and their loci in the LGGS fields. Confirmed GCs are
  marked with circles, GC candidates are denoted as triangles, while
  other objects are indicated by pluses. The large ellipse is the
  $D_{25}$ boundary of the M31 disk \citep{rac91}, while the two small
  ellipses are the $D_{25}$ contours of NGC 205 (northwest) and M32
  (southeast). The ten large squares are the LGGS fields.}
  \label{fig1}
\end{figure}

We selected our sample clusters from the RBC v.4.0, which is a
compilation of photometry and identifications from many previous
catalogs.
We used archival $UBVRI$ images from the LGGS, which covers a region
of 2.2 deg$^2$ along the galaxy's major axis. The images we used
consisted of 10 separate but overlapping fields with a scale from
0.261\arcs pixel$^{-1}$ at the center to 0.258\arcs pixel$^{-1}$ in
the corners of each image. The field of view of each mosaic image is
$36'\times36'$. The observations were taken from August 2000 to
September 2002 with the KPNO 4 m telescope. The median seeing of the
LGGS images is $\sim$ 1\arcs. \citet{cald09} inspected the images and
found some new clusters, including over one hundred young clusters. To
limit the scope of our work, we only perform photometry of the
clusters in the LGGS images using the identifications provided by
\citet{cald09}. We employed {\sc iraf/daofind} to find the sources in
the images and match them to the RBC coordinates. We matched 1191
objects in all LGGS fields in the $V$ band. To prevent mistakes, we
checked each object visually in the images. Like \citet{cald09},
we also ran into positional errors for five objects (NB18, B353, NB60,
V229, and NB57) in the RBC v.4.0; these five sources were not detected
at the RBC coordinates in the LGGS images. In addition, 138 clusters
listed in the RBC v.4.0 are actually composed of two or more
individual stars in the images. Another five objects (BH18, B088D,
B091D, C011, and B117) have very bright stars nearby, causing
contamination; 32 are saturated, ten are too faint to be detected, 18
have very strong background gradients and one nebula (SK150C), four
(SK156B, SK099A, SK062B, and SK105B) have very high backgrounds, and
eight are suspected to be galaxies. In total, we excluded those 221
objects to make our sample as clean as possible. Eventually, we
detected 970 RBC objects (including all object types) in the LGGS
survey images, for which we will perform photometry in this
paper. Fig. \ref{fig1} shows the spatial distribution of the 970
objects in the LGGS fields. Confirmed GCs are marked with circles, GC
candidates are denoted as triangles, while other objects are indicated
by pluses. The large ellipse is the $D_{25}$ boundary of the M31 disk
\citep{rac91}, while the two small ellipses are the $D_{25}$ contours
of NGC 205 (northwest) and M32 (southeast). The ten large squares are
the LGGS field boundaries. We will mainly focus on the confirmed and
candidate globular-like clusters ($f=1$ and 2 in the RBC v.4.0,
respectively), since we anticipate the presence of a significant
fraction of YMCs in this sample.

\subsection{Integrated photometry}
\label{s:phot}

We used the LGGS archival images of M31 in the $UBVRI$ bands to
perform photometry. Previously, \citet{massey} compiled
point-spread-function (PSF) photometry for 371,718 stars (point
sources) in the M31 fields, with photometric uncertainties of $<10$\%
below $V = 23$ mag. However, there is as yet no published LGGS
photometry for extended sources, such as star clusters and
galaxies. Recently, \citet{cald09} undertook aperture photometry of the
resolved star clusters only in the $V$ band and studied the nature of
over one hundred young M31 clusters. However, LGGS photometry of M31
clusters in the other bands ($BVRI$) has not yet been compiled.

For this reason, we perform aperture photometry of the M31 clusters
found in the LGGS images in all of the $UBVRI$ bands to provide a
comprehensive and homogeneous photometric catalog of M31 globular-like
clusters. The photometry routine we used is {\sc iraf/daophot}
\citep{stet}. Following \citet{cald09}, we use eight different
aperture sizes (with radii of $r_{\rm ap} = 1.03, 1.64, 2.19, 2.90,
3.86, 5.13, 6.82$, and 9.06\arcs) to ensure that we adopt the most
appropriate photometric radius that includes all light from the
objects, but excludes (as much as possible and to the extent that this
was obvious) extraneous field stars. We decided on the size of the
aperture needed for the photometry based on visual examination. The
local sky background was measured in an annulus with an inner radius
of 9.29\arcs and 2.58\arcs wide. (Although we performed our cluster
photometry using different apertures, we chose to use identical
background annuli for convenience. We tested the validity of this
approach for one source based on using different apertures and
different backgrounds, including background gradients.  The results
show that background variations result in uncertainties at a level of
only $\sim0.001$ mag.) To check whether and how seriously
aperture variations affect our results, we performed tests with a
series of different apertures, ranging from the proper radius given in
Table~\ref{photo} to a radius of 10\arcs larger than the tabulated
value. As a result, the magnitudes typically only vary by $\sim0.06$
mag due to background variations and contamination from other
sources. The instrumental magnitudes were then calibrated to the
standard Johnson-Kron-Cousins $UBVRI$ system by comparing the
published magnitudes of stars from \citet{massey}, who calibrated
their photometry with standard stars of \citet{lan92}, with our
instrumental magnitudes. Since the magnitudes in \citet{massey} are
given in the Vega system, our photometry is also tied to that
system. Finally, we obtained photometry for 970 objects, with 965,
967, 965, 953, and 827 sources in the individual $UBVRI$ bands,
respectively. After matching our photometry with the measurements
in the RBC v.4.0, \citet{barmby}, and \citet{peac10}, we found that
205, 123, 14, 126, and 109 objects (in the corresponding passbands) do
not have any previously published photometry. We remind the reader
that the photometry of \citet{peac10} is in the $ugriz$ system, which
we transferred to the Johnson-Kron-Cousins $UBVRI$ system based on
\citet{jester}. (Their equations were derived for a sample of stars
with $R-I<1.15$ mag.) Table \ref{photo} lists our new $UBVRI$
magnitudes and the aperture radii used. The table only includes
photometry for objects that are located in the LGGS images, with
errors given by {\sc iraf/daophot}.  The object names follow the
naming convention of \citet{barmby}, \citet{perrett}, and
\citet{gall04,gall06,gall07}.

\begin{figure*}
\centerline{
\includegraphics[scale=.5,angle=-90]{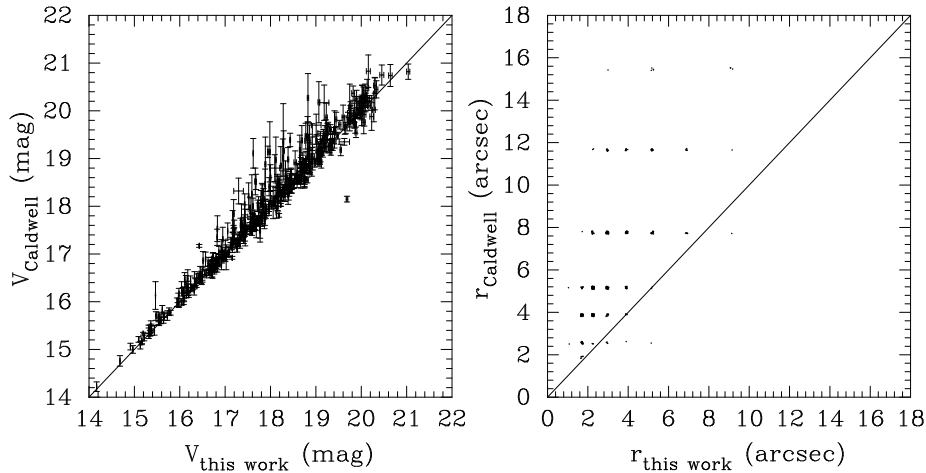}}
\caption[]{{\it(left)} Comparison of our $V$-band photometry with
  previous measurements from \citet{cald09}. {\it(right)} Comparison
  of the corresponding apertures for photometry. Point sizes represent
  the frequency of the apertures used.}
  \label{fig2}
\end{figure*}

Since we used the same images as \citet{cald09}, a direct
comparison can tell us whether or not our photometry is
reliable. However, \citet{cald09} only include photometry in the $V$
band. Figure~\ref{fig2} shows comparisons of the $V$-band photometry
and apertures. We note that in the left-hand panel, the $V$ band
magnitudes are very consistent. In the right-hand panel, we use the
point size to represent the frequency of the aperture used. It is
clear that for small apertures the two methods are reasonably
consistent. However, for larger apertures, our apertures are a few
arcseconds smaller than \citet{cald09}'s. In fact, the aperture of
every source considered here was determined by visual checks to make
sure that it was large enough, but not too large (to avoid
contamination from other sources). The left-hand panel proves that,
with a small number of exceptions, the choice of aperture does not
significantly affect the resulting photometry.

\begin{figure}
\centerline{
\includegraphics[scale=.45,angle=-90]{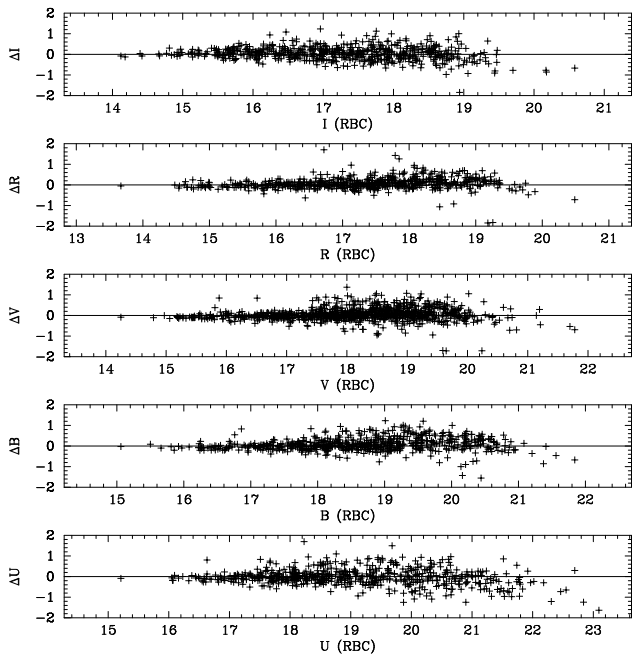}}
\caption[]{Comparisons of our photometry of 970 objects in the $UBVRI$
  bands with previous measurements from the RBC v.4.0. Note the good
  overall agreement in all filters. $\Delta$mag = our measurement $-$
  literature value.}
  \label{fig3}
\end{figure}

To examine the quality and reliability of our photometry, we show
comparisons of the aperture magnitudes of the 970 objects considered
here with the magnitudes collected from various sources in the RBC
v.4.0 in Fig. \ref{fig3}. We find good agreement in all bands, with an
rms scatter in the photometric differences (throughout this paper
defined in the sense of our determination minus literature
measurement) ranging from $\sigma=0.325$ mag in the $I$ band (590
objects) to $\sigma=0.398$ in the $U$ band (586 objects). This is in
spite of the fact that no effort was made to ensure that the apertures
used in both data sets were the same. The photometric offsets (defined
as $\sigma/\sqrt{N}$, where $\sigma$ is the standard deviation and $N$
the number of data points) and the rms scatter of the differences
between the RBC v4.0 set and our new magnitudes are summarized in
Table \ref{photocom}, showing no apparent systematic uncertainties.
(The offsets are $\sim0.1$ to $0.2\sigma$. In the $U$ and $BVRI$
filters, our photometry is brighter and fainter, respectively, at
these levels than the RBC v.4.0 compilation.)

\begin{figure}
\centerline{
\includegraphics[scale=.45,angle=-90]{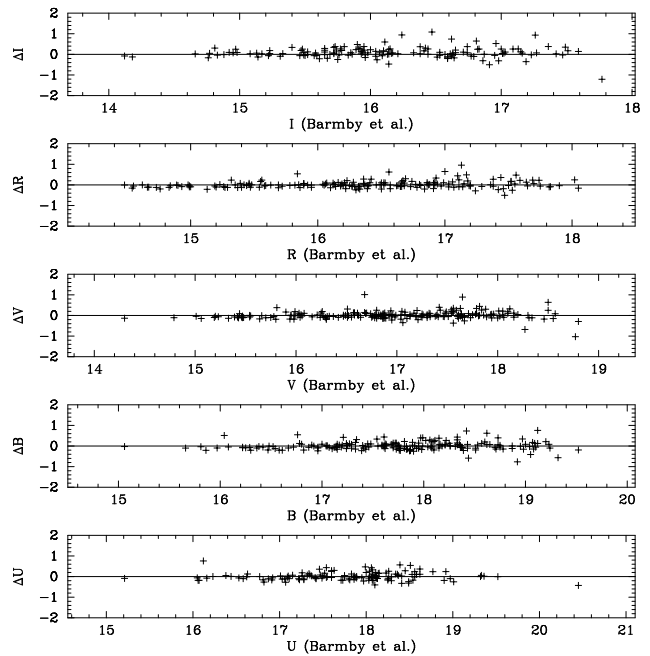}}
\caption[]{Comparisons of our photometry with previous measurements
  from \citet{barmby}. Note that the scatter is smaller in all filters
  than in Fig. \ref{fig3}. $\Delta$mag is defined as in
  Fig. \ref{fig3}.}
  \label{fig4}
\end{figure}

We also compare our photometry with previous measurements from
\citet{barmby}, which were based on one photometric system and much
more homogeneous in nature than the RBC v.4.0 data. Fig.~\ref{fig4}
shows the comparison. The photometric offsets and rms scatter of the
differences between their and our magnitudes are summarized in Table
\ref{photocomb}.

\begin{figure}
\centerline{
\includegraphics[scale=.45,angle=-90]{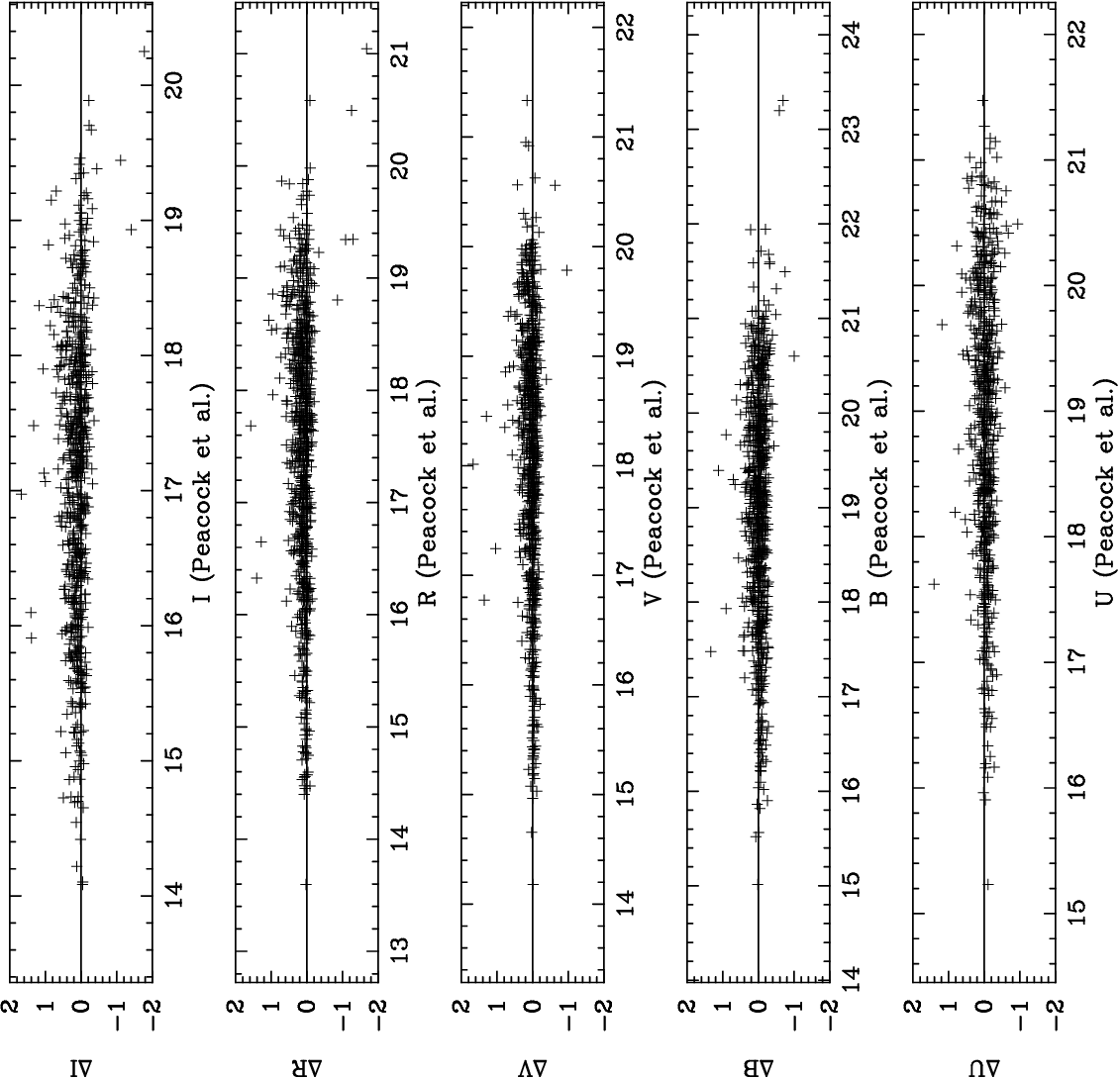}}
\caption[]{Comparisons of our photometry with previous measurements
  from \citet{peac10}. Note that the scatter is smaller in all filters
  than in Fig. \ref{fig3}. $\Delta$mag is defined as in
  Fig. \ref{fig3}.}
  \label{fig5}
\end{figure}

We also compared our photometry with previous measurements from
\citet{peac10}, whose photometry is based on the SDSS $ugriz$ system
(which is much more homogeneous in nature than the RBC v.4.0 data). As
discussed in \S\ref{s:phot}, we transferred the SDSS $ugriz$ system to
the standard broadband system and show the comparison in
Fig.~\ref{fig5}. As for Table~\ref{photocomb}, the photometric offsets
and rms scatter of the differences between their and our magnitudes
are summarized in Table \ref{photocomp}.

\begin{figure*}
\centerline{
\includegraphics[scale=.55,angle=0]{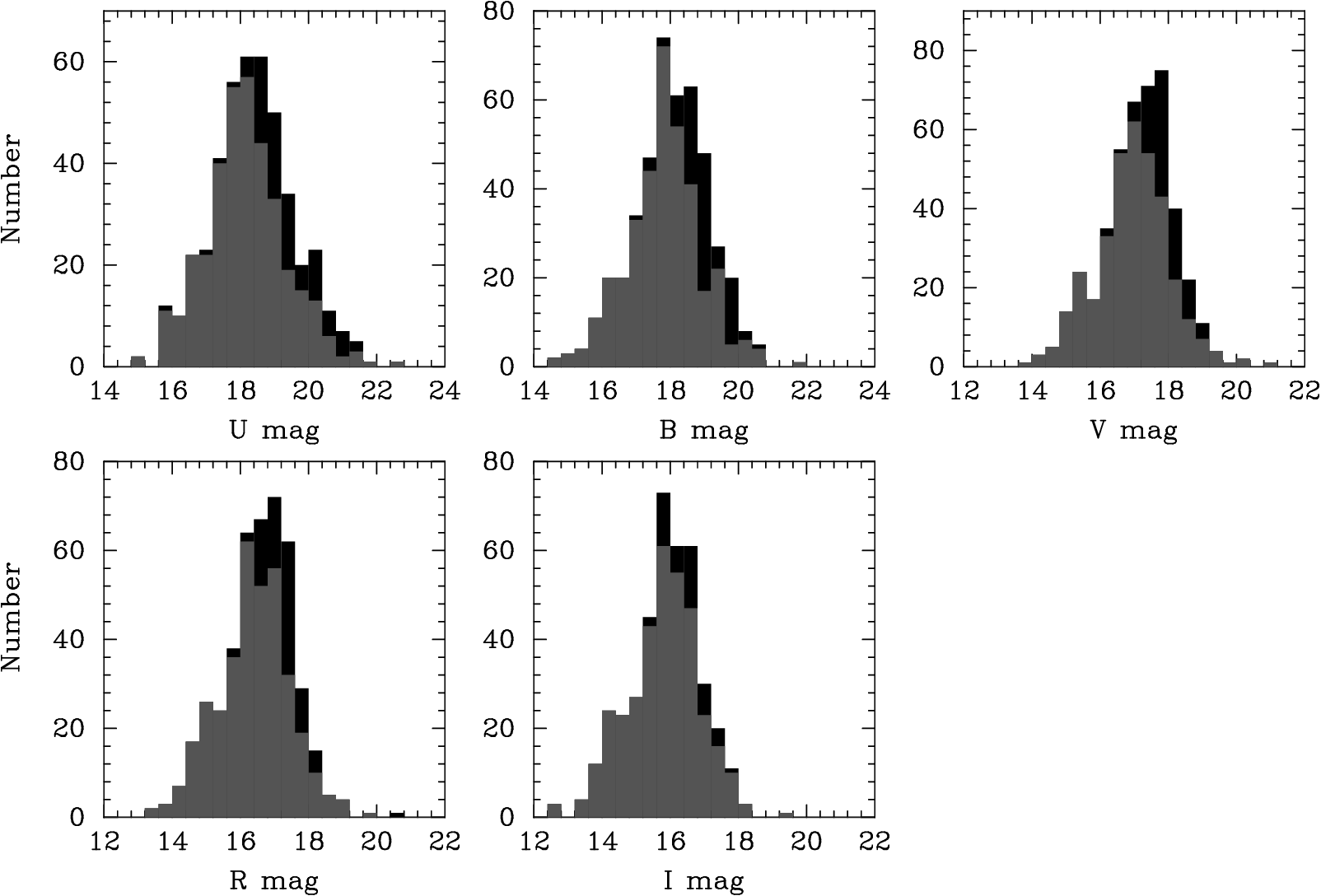}}
\caption[]{$UBVRI$ distributions of the globular-like star clusters
  and candidates in M31 (see Table~\ref{photo}).  The black filled
  histograms contain both the confirmed clusters and candidates ($f=1$
  and 2 in the RBC), while the gray histograms are subsamples that are
  composed of only confirmed clusters ($f=1$).}
  \label{fig6}
\end{figure*}

Fig. \ref{fig6} shows the distributions of the $UBVRI$ magnitudes of
the M31 star clusters from Table~\ref{photo}. The black filled
histograms contain the full sample of confirmed, globular-like star
clusters ($f=1$ in the RBC) and candidates ($f=2$), while the gray
filled histograms include only the confirmed globular-like
clusters. The cluster candidates are mostly found near the faint ends
of the distributions, as expected. The peaks of the distributions in
all bands agree well with those of the confirmed M31 GCs reported by
\citet{gall06} and \citet{fan09}, suggesting that our selection of
confirmed, globular-like star clusters is not systematically biased.

\section{Results and analysis}
\label{s:ana}

In this section, we describe the methods and processes used for the
determination of the cluster ages and masses based on SED fitting.
We supplemented our photometry (Table \ref{photo}) with RBC v.4.0
measurements if we could not obtain the relevant measurements
ourselves from the LGGS images.\footnote{We obtained photometry for
970 objects from the LGGS images, which we combined with RBC v.4.0
magnitudes (2045 objects) to compile the largest possible
photometric sample. We selected the $f=1$ and 2 clusters with
photometry in no fewer than six of the eight available filters
($UBVRIJKH$) for our age determinations, yielding 445 objects. A
large number of these 445 clusters are located in the galaxy's halo
(see Figs. \ref{fig14} and \ref{fig15}) and not covered by the LGGS
images (see Fig. \ref{fig1}). For most of these halo clusters, the
photometry was therefore completely obtained from the RBC.} This
should not introduce any additional systematic effects, given the
absence of large systematic offsets ($>0.1$ mag) in Figs \ref{fig3},
\ref{fig4} and \ref{fig5}. Our final working sample is composed of
445 confirmed globular-like clusters and candidates. It is important
to keep in mind that the sample also includes a small number of
clusters with photometry completely taken from the literature. Our
aim of this selection procedure is to base our results on the
largest possible cluster sample.

\subsection{Ages of the globular-like clusters}
\label{s:age}

From an observational point of view, studying M31 star clusters is
complicated, since in most cases we only have access to their
integrated spectra and photometry and cannot study the resolved
stellar populations. Therefore, we can only obtain their key physical
parameters, such as ages and metallicities, by careful analysis of the
integrated observables. However, a large body of evidence suggests
that a strong age-metallicity degeneracy dominates if only optical
photometry is used \citep{wor94,ar96,kaviraj07}. \citet{anders04a}
studied the star clusters in NGC 1569 using multiwavelength {\sl HST}
observations. They strongly recommend to use NIR photometry as the
only way to break the degeneracy for young clusters \citep[see
also][]{jong96}. \citet{anders04b} investigated the systematic
uncertainties inherent to SEDs fits in the $UBVRIJH$ bands based on
stellar population synthesis modeling and found that access to at
least one NIR passband can significantly improve the results and
constrain the metallicity. They concluded that the degeneracy can be
partially broken by adding NIR photometry to the optical colors,
depending on the age of the stellar population. \citet{deg05a} and
\citet{wu05} also showed that the use of NIR colors can greatly
contribute to break the age-metallicity and age-extinction
degeneracies. Thus, in our fits, we will combine our $UBVRI$
photometry with $JHK$ photometry from the RBC v.4.0 to disentangle the
degeneracies and obtain more accurate results.

The RBC contains 703 confirmed or candidates M31 GCs that have
homogeneous NIR $JHK$ data, of which the majority were obtained from
2MASS photometry by \citet{gall04} and only 17 are from previously
published data. For point sources, the authors use $r=4$ arcsec while
for extended sources they use $r=5$ arcsec. Further, \citet{gall04}
show that their 2MASS photometry is quite consistent with pre-2MASS
photometry. As for the NIR $JHK$ magnitudes, the uncertainties in the
fitting routines are estimated as in \citet{fan06} by applying the
relations in figure~2 of \citet{chs01}, which shows the observed
magnitude uncertainty as a function of magnitude for bright stars in
the 2MASS $JHK$ bands.

We use a $\chi^2$ minimization technique for our age estimates,
comparing the observed integrated SEDs with theoretical SSP models. We
take advantage of the BC03 SSP models, using Padova 1994 evolutionary
tracks and a \citet{chab} initial mass function (IMF) with lower and
upper mass cutoffs of 0.1 and $100 M_\odot$, respectively. The BC03
SSP synthesis models include six initial metallicities, $Z=0.0001,
0.0004, 0.004, 0.008, 0.02$ (solar), and 0.05, with 221 unequally
spaced time steps from 0 to 20 Gyr. Following \citet{fan06},
\citet{ma07,ma09}, and \citet{wang10}, but improving on their
approaches, a new, higher-resolution spectral grid containing 100
metallicities (from $Z=0.0001$ to 0.05) was created by interpolating
in logarithmic space, with equally spaced intervals of log $Z$ between
the newly created templates.

The BC03 SSP model spectra can be easily convolved to magnitudes in
the AB system using the filter-response functions in the $UBVRIJHK$
bands. The apparent magnitudes of the BC03 SSP synthesis models in the
AB system are given by
\begin{equation}
m_{\rm AB}(t)=-2.5~{\log~\frac{\int_{\lambda_1}^{\lambda_2}{{\rm
        d}\lambda}~
    {\lambda}~F_{\lambda}(\lambda,t)~R(\lambda)}{\int_{\lambda_{1}}^{\lambda_{2}}{{\rm
        d} \lambda}~{\lambda}~R(\lambda)}}-48.60,
\label{eq1}
\end{equation}
where $R(\lambda)$ is filter-response function and
$F_{\lambda}(\lambda,t)$ is the flux, which is a function of
wavelength ($\lambda$) and evolutionary time ($t$). $\lambda_1$ and
$\lambda_2$ are the lower and upper wavelength cutoffs of the
respective filter (see BC03).

Since all our photometric measurements ($UBVRIJHK$, both our own
photometry and that from the RBC) are calibrated in the Vega system,
for convenience of comparison we need to convert the observed
integrated magnitudes to the AB system using the \citet{kuru} SEDs.

For those clusters in our sample that have available reddening
values from \citet{fan08} or \citet{barmby}, the magnitudes were
corrected for reddening assuming a \citet{ccm} extinction curve, so
that their ages ($t$) can be determined by comparing the interpolated
high-resolution BC03 SSP synthesis models with the SEDs from our
photometry and with $Z$ as a free parameter, i.e.,
\begin{equation}
\chi^2_{\rm min}(t,Z)={\rm
  min}\left[\sum_{i=1}^8\left({\frac{m_{\lambda_i}^{\rm
        obs}-m_{\lambda_i}^{\rm mod}}{\sigma_i}}\right)^2\right],
\label{eq2}
\end{equation}
where $m_{\lambda_i}^{\rm mod}(t,Z)$ is the integrated magnitude in
the $i^{\rm th}$ filter in a theoretical SSP at age $t$ and for
metallicity $Z$, $m_{\lambda_i}^{\rm obs}$ represents the observed,
integrated magnitude in the same filter, $m_{\lambda_i}=UBVRIJHK$, and
\begin{equation}
\sigma_i^{2}=\sigma_{{\rm obs},i}^{2}+\sigma_{{\rm mod},i}^{2}.
\label{eq3}
\end{equation}
Here, $\sigma_{{\rm obs},i}$ is the observational uncertainty. 
Since the RBC does not include any magnitude uncertainties, we applied
the rough estimates from \citet{gall04}, i.e., 0.05 and 0.08 mag for
the $BVRI$ and $U$ bands, respectively. As for the NIR $JHK$
magnitudes, the uncertainties are estimated as in \citet{fan06} by
applying the relations in figure~2 of \citet{chs01}, which shows the
observed uncertainty as a function of magnitude for bright stars in
the 2MASS $JHK$ bands. In addition, \citet{fan06} proved that the
adopted uncertainty does not affect the SED fits. $\sigma_{{\rm
mod},i}$ represents the uncertainty associated with the model itself,
for the $i^{\rm th}$ filter. Following \citet{deg05a}, \citet{wu05},
\citet{fan06}, \citet{ma07,ma09}, and \citet{wang10}, we adopt
$\sigma_{{\rm mod},i}=0.05$.

For clusters without reddening values from the literature, we
constrained the ages while keeping $Z$ and reddening as free
parameters, using
\begin{equation}
\chi^2_{\rm min}[t,Z,E(B-V)]={\rm
  min}\left[\sum_{i=1}^8\left({\frac{m_{\lambda_i}^{\rm
        obs}-m_{\lambda_i}^{\rm mod}}{\sigma_i}}\right)^2\right].
\label{eq4}
\end{equation}
We varied the reddening between $E(B-V)=0.0$ and 2.0 mag in steps of
0.02 mag.

\begin{figure*}
\centerline{
\includegraphics[scale=0.55,angle=-90]{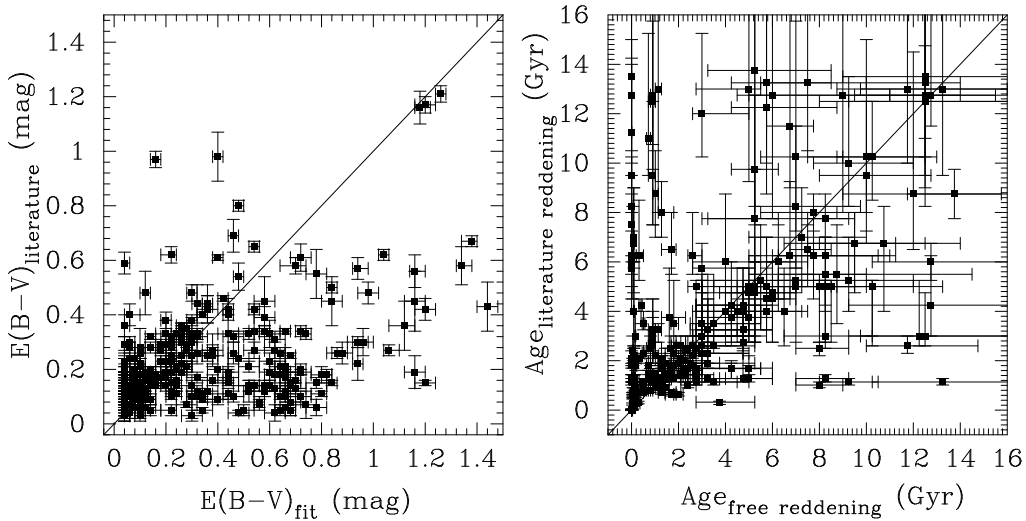}}
\caption[]{{\it (left)} Comparison of reddening from the literature
and from adopting reddening as a free parameter, both based on
Eq. (\ref{eq4}). (right) Comparison of ages obtained with the
reddening values from the literature and those fitted adopting
reddening as a free parameter.}
  \label{fig7}
\end{figure*}

To check the consistency of our reddening estimates with respect to
those of \citet{fan08} and \citet{barmby}, we refitted our cluster
SEDs adopting reddening as a free parameter in Eq. (\ref{eq4}). Fig.
\ref{fig7} shows the comparison. We note that the scatter is large,
although there is a tendency that our fits with reddening as a free
parameter resulted in higher extinction estimates than the values from
the literature. In turn, this affects the resulting age estimates
through the age-extinction degeneracy: see the right-hand panel of
Fig. \ref{fig7}, which shows a slight tendency toward younger ages
resulting from our `free-reddening' fits than when using the smaller
extinction values from the literature (the effect is small because the
differences in our extinction estimates are not very large either).
It is notoriously difficult to obtain reliable reddening estimates
from broad-band fits, independent of the wavelength range covered
\citep{anders04b,deg05a,da06}. \citet{deg05a} and \citet{da06}
compared the systematic differences resulting from using different
approaches based on broad-band photometry and concluded that the
differences are very small. The right-hand panel of Fig. \ref{fig7}
provides some support for their earlier conclusion.

\begin{figure}
\centerline{
\includegraphics[scale=0.38,angle=-90]{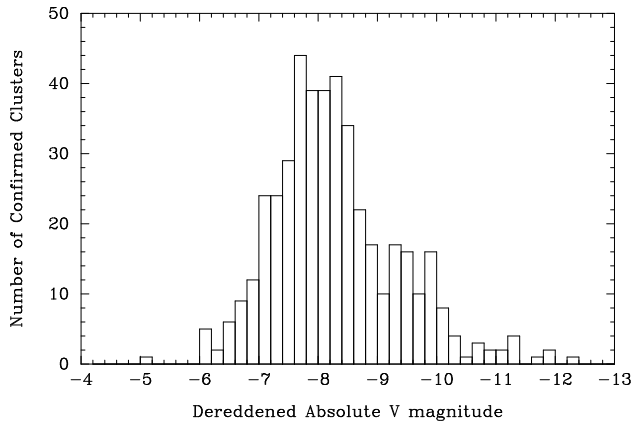}}
\caption[]{Absolute, extinction-corrected $V$-band magnitude
  distribution of the confirmed M31 star clusters in our sample.}
  \label{fig8}
\end{figure}

To get a rough handle on the completeness of our photometry, we plot
the distribution of absolute (extinction-corrected) $V$ magnitudes of
our 445 sample objects in Fig.~\ref{fig8}. Using a bin size of 0.2
mag, our proxy for the overall limiting magnitude is $M_V=-7$ mag
(i.e., the half-peak point of the distribution). In fact,
\citet{massey} claimed that the survey reaches $UBVRI \sim23$ mag,
corresponding to $M_V=-1.47$ mag if we adopt a distance modulus of
$(m-M)_0=24.47$ mag \citep{mc05}, with $< $10\% uncertainty.

If the initially estimated age of a star cluster is older than 14 Gyr,
we adopt an age of 12 Gyr and iterate until the fitting routine
reaches a local minimum. It is well-known that SSP SEDs are not
sensitive to changes in age for ages $>10$ Gyr
\citep[e.g.,][]{ma07}. Therefore, although the upper age limit in the
BC03 models is 20 Gyr, the ages of the clusters determined here do not
exceed 14 Gyr. The estimated ages of the M31 globular-like clusters in
our sample are listed in Table~\ref{agemass}. We also provide the
$1\sigma$ errors. When fitting a given parameter, we first fix all
other parameters to their best values and fit the minimum $\chi^2$. We
subsequently vary the parameter of interest and record an error
corresponding to the $1\sigma \chi^2$ value.

\begin{figure*}
\centerline{
\includegraphics[scale=0.5,angle=-90]{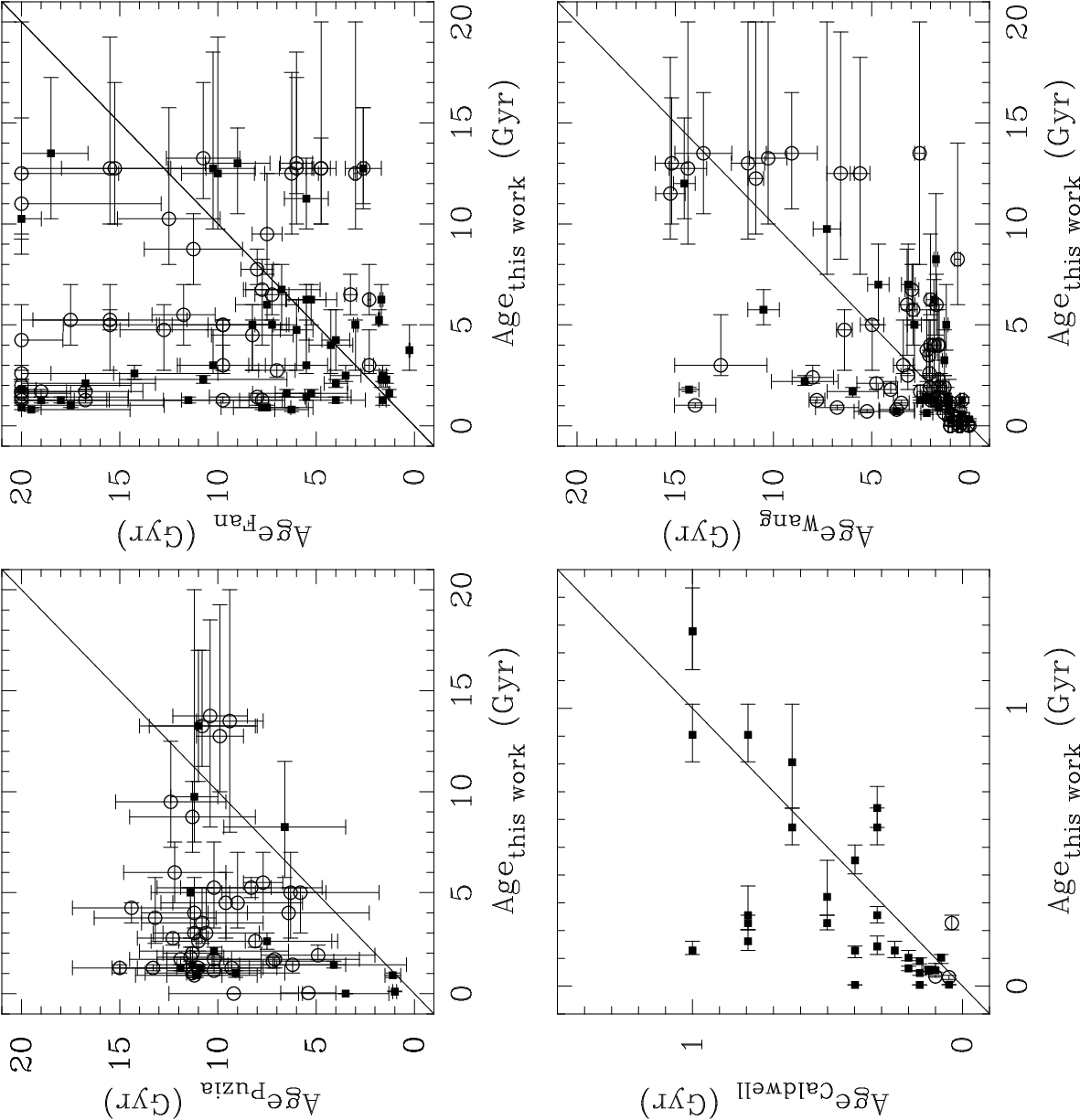}}
\caption[]{Comparisons of the age determinations of the 445 confirmed
  clusters and candidates in our sample in common with (top left)
  \citet{puzia05}, (top right) \citet{fan06}, (bottom left)
  \citet{cald09}, and (bottom right) \citet{wang10}. The black
  squares represent clusters whose ages we constrained based on our
  new photometry (combined with 2MASS data), while the circles are
  clusters whose ages were constrained using combined photometry from
  this paper, the RBC v.4.0, and 2MASS.}
  \label{fig9}
\end{figure*}

Fig. \ref{fig9} shows comparisons of our newly derived ages with
previous results. The top left-hand panel is a comparison between
\citet{puzia05} and our results. We note that \citet{puzia05}'s
cluster ages are systematically older than ours. This may have been
caused by their assumption of 13 Gyr as initial guess for their age
iterations, possibly leading to convergence in a different local
minimum. (We did not constrain the initial guesses.) The top
right-hand panel compares the results of \citet{fan06} with our new
determinations. We note that the scatter is large, but so are the
realistic error bars. Any systematic differences, particularly for
younger objects, may be due to the age-metallicity degeneracy
affecting broad-band SED fitting: we previously adopted metallicities
deemed suitable for the old M31 GCs. We show in Fig. \ref{fig11} that
this is indeed the case: the metallicity determinations of
\citet{perrett}, who used Galactic GCs to calibrate their metallicity
scale, are systematically more metal poor than those derived from our
SED fits. This could lead to the systematically older ages of
\citet{fan06}. The bottom left-hand panel is a comparison between our
new results and those from \citet{cald09}, who only estimated the ages
of their young clusters. The agreement is good. The bottom right-hand
panel compares our new age estimates with those of \citet{wang10}, who
applied the GALEV SSP models for their SED fits. Both sets of results
are consistent, especially for clusters younger than 2 Gyr and within
the fairly large uncertainties \citep[cf.][]{deg05a}.

The spectroscopic age estimates are important for comparison with
our work. \citet{cald09} estimated the ages of 134 young and 367 old
clusters. In our sample, the fraction of clusters that are classified
by \citet{cald09} as old/young is 314/29, while the fraction for the
same clusters based on our new age determinations is 181/162.
\citet{cald09} did not provide the uncertainties associated with their
age estimates. However, if we consider the lower (upper) limits to the
ages derived here, the fraction is 169/174 (194/149). In our full
sample, this fraction is 224/221. Again, if we consider the
uncertainties in our age estimates, we arrive at respective fractions
of 210/235 and 244/201. Even if we were to use our reddening-free
fitting results, the fraction does not change by much, which means
that the reddening values from the literature that we applied in our
work do not affect our results significantly. The difference is
essentially due to the different methods applied. The similar study of
\citet{wang10} also shows that the fraction of old/young clusters is
low. In addition, the proportion of young clusters in the disk should
not be very small, while we also note that a large fraction of our
sample should be composed of young massive clusters, rather than
typical GCs.

\begin{figure}
\centerline{
\includegraphics[scale=0.45,angle=-90]{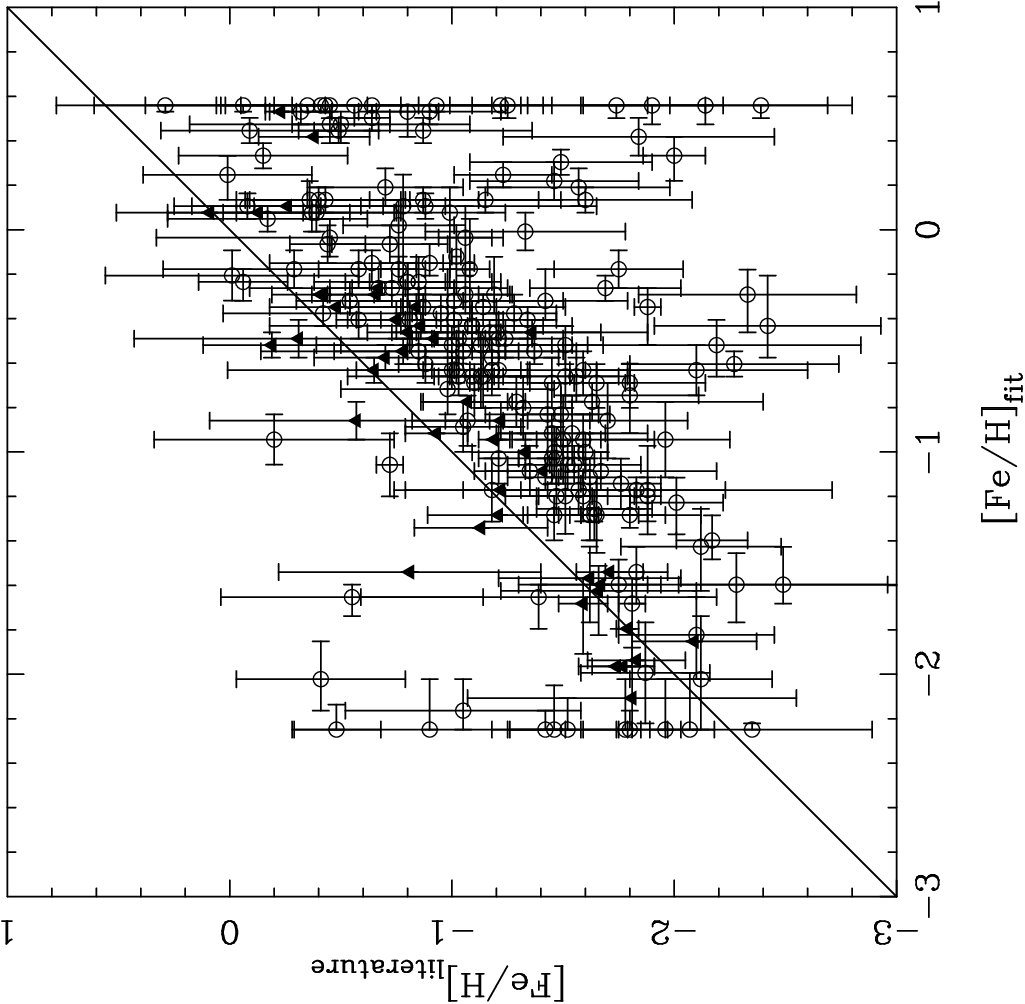}}
\caption[]{Comparison of the metallicities derived from our SED fits
  with those from \citet{perrett}, \citet{barmby}, and
  \citet{hbk91}. The latter are systematically more metal poor than
  those derived from SED fitting, which could lead to the
  systematically older ages of \citet{fan06}. The filled triangles
  represent old clusters with ages $>10$ Gyr, while the circles are
  young clusters with ages $<10$ Gyr from our estimates. Most of the
  clusters are young, which may cause the metallicity bias.}
  \label{fig10}
\end{figure}

\begin{figure}
\centerline{
\includegraphics[scale=0.45,angle=-90]{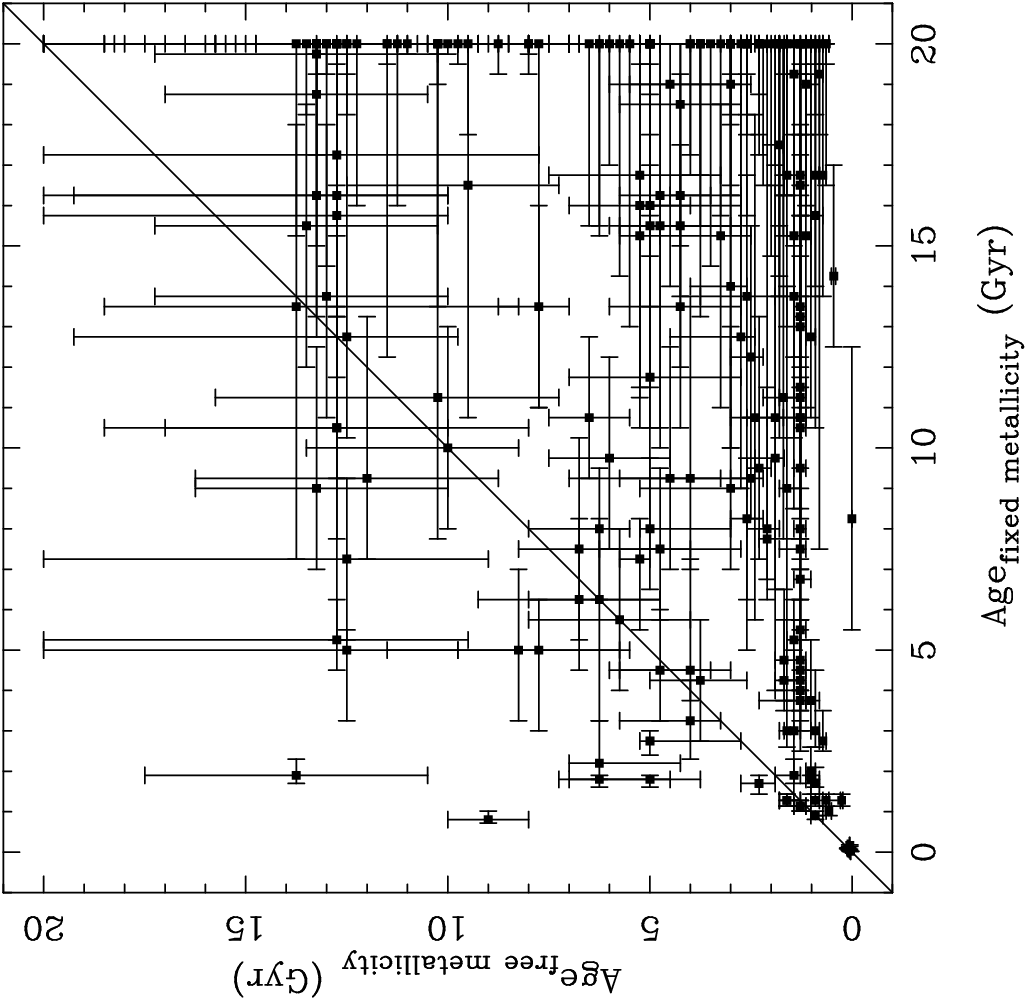}}
\caption[]{Comparison of the ages fitted with free metallicity and
  those with metallicity from \citet{perrett}, \citet{barmby} and
  \citet{hbk91}. The reddening values of the two methods are both from
  \citet{fan08}.}
  \label{fig11}
\end{figure}

To check our results, we need to compare the metallicities
derived from our SED fits with literature values based on
spectroscopy. We applied the metallicity estimates from Table 1 of
\citet{fan08}, which contains a summary of the spectroscopic
metallicities from \citet{perrett}, \citet{barmby}, and \citet{hbk91},
all of which were calibrated using old Galactic GCs.  In addition,
\citet{fan08} showed that there are no systematic offsets between
these three sets of measurements and that we can use them
safely. Figure \ref{fig10} shows a comparison of the metallicities
derived from our SED fits with those from the literature. The latter
are systematically more metal poor than those derived from SED
fitting, which could lead to the systematically older ages of
\citet{fan06}. As \citet{fp05} pointed out, many of the `old' clusters
from \citet{perrett} are actually young, since they used metallicities
that were calibrated based on old Galactic GCs. Thus, the
metallicities of such clusters should be more metal rich. This may
well explain the apparent metallicity bias in Fig. \ref{fig10}.

We also compared the ages obtained with the metallicity estimates
from \citet{perrett}, \citet{barmby}, and \citet{hbk91} with those
obtained based on free-metallicity fits (see Fig.~\ref{fig11}). The
reddening values used were fixed at the values of \citet{fan08} and
\citet{barmby}, as before. A large number of young clusters in the
free-metallicity fits are considered old using the fixed-metallicity
fits based on literature values, which may be due to the
age-metallicity degeneracy. If these young clusters in
\citet{perrett}, \citet{barmby}, and \citet{hbk91} were considered
old, their metallicity would be poorer than the real value, which can
also explain the bias in Fig. \ref{fig10}. These two figures show the
effects of the age-metallicity degeneracy very clearly.

\begin{figure}
\centerline{
\includegraphics[scale=0.33,angle=-90]{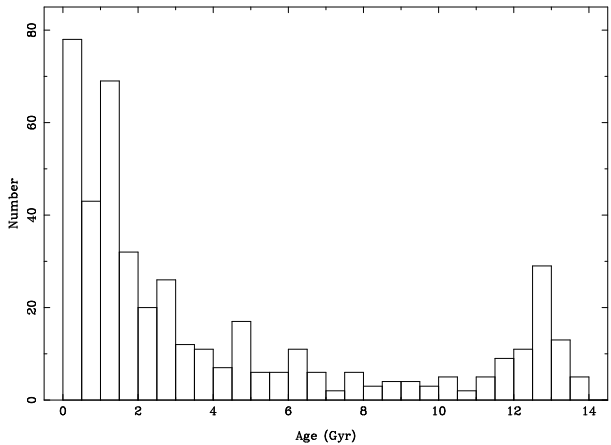}}
\caption[]{Age distribution of GCs in M31. Of these, 121 of the 445
  sample objects are younger than 1 Gyr ($\sim 27$\%) and 221 are
  younger than 2 Gyr ($\sim 50$\%), suggesting that about half of the
  M31 globular-like clusters formed during the past 2 Gyr, which
  is a quite similar result to that of \citet{wang10}.}
  \label{fig12}
\end{figure}

The cluster age distribution is interesting because it offers a clue
to the galaxy's formation history. Fig. \ref{fig12} shows the age
distribution of our sample of globular-like clusters in M31 (bin size:
0.5 Gyr), which is very similar to that of \citet{wang10} who
obtained the ages of M31 GCs based on a similar SED-fitting
method. In the entire sample, 121 and 221 of the 445 star clusters
are younger than 1 and 2 Gyr, respectively, corresponding to $\sim$27
and 50\% of the sample \citep[see
also][]{will01a,will01b,bur04,beasley,puzia05,fan06,ma09,cald09,
wang10,per10}. The age distribution of our clusters, combined with
those of the younger clusters from previous studies, shows evidence of
active star formation in M31 over the past 2 Gyr. This implies that
there may have been several star-forming episodes in this period,
possibly triggered by (a) major or several minor mergers with other
galaxies (see below). In addition, we also find some intermediate-age
star clusters, with ages between 3 and 8 Gyr. These
intermediate-age counterparts have also been found by \citet{puzia05},
\citet{fan06}, \citet{ma09}, and \citet{wang10}, but there are some
differences between our work and these previous studies. For instance,
the intermediate-age ($\sim 3 - 8$ Gyr-old) cluster population in
\citet{puzia05} is found to be $\sim 2 - 5$ Gyr in this paper, while
our newly determined intermediate-age clusters range from 2 to 20 Gyr
in \citet{fan06} and from 1 to 10 Gyr in \citet{wang10}. This is most
likely due to the different models and fitting methods applied
\citep[cf.][]{deg05a}. The age distribution of the M31 globular-like
clusters is quite different from that of the Milky Way GCs, which are
all older than 10 Gyr. In fact, this might be due to our sample
selection. Our results are based on observations in the M31 disk,
where most of the young clusters are located. We also note that our
final sample includes 445 objects that are located in the disk. These
disk clusters are usually young and this will cause the distribution
to be dominated by the young population.  \citet{hmm07} pointed out
that the Milky Way has had an exceptionally quiet formation history
over the last 10 Gyr, which might explain the lack of a significant
population of young globular-like clusters in our Galaxy. It may also
explain why M31 and the Milky Way have similar sizes, masses, and
Hubble types, but M31 has many more GCs \citep[$460\pm70$,][]{bh01}
than the Milky Way \citep[$>$152,][]{fms07}, i.e., this is probably
because M31 underwent a much more extended star-formation history than
the Milky Way. \citet{iba05} and \citet{hmm07} suggested that a
recent, active merger may have occurred in M31, which could have
triggered GC formation in the interval from $\sim$8 to less than 1 Gyr
ago.  \citet{mc09} suggested that an encounter between M33 and M31
took place a few Gyr ago. The clusters with ages in excess of 10 Gyr
in Fig. \ref{fig12} were most likely created when the galaxy formed,
while the young globular-like clusters might have been created in a
number of mergers during the last few Gyr or by the recent galactic
encounter with M33.

\begin{figure}
\centerline{
\includegraphics[scale=0.33,angle=-90]{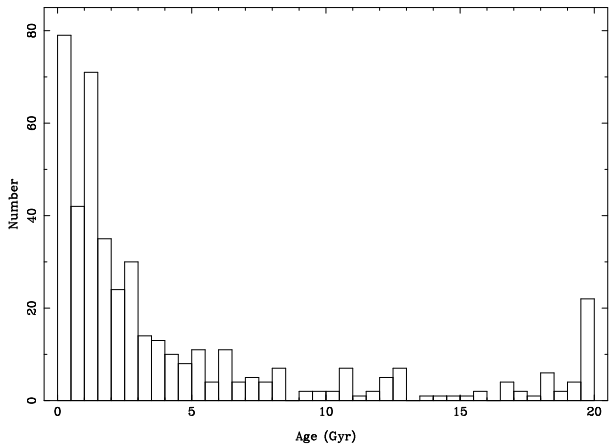}}
\caption[]{Age distribution of GCs in M31. Of these, 121 of the 445
  sample objects are younger than 1 Gyr ($\sim 27$\%) and 221 are
  younger than 2 Gyr ($\sim 50$\%), suggesting that about half of the
  M31 globular-like clusters formed during the past 2 Gyr.}
  \label{fig13}
\end{figure}

The peak around 12.5 Gyr in Fig.~\ref{fig12} is artificially
created by our method of resetting the fits for ages exceeding 14
Gyr. To show this, we plot the age distribution resulting from
allowing the model to reach up to 20 Gyr in Fig.~\ref{fig13}.

We converted the clusters' angular positions to a linear scale at the
distance of M31, 785 kpc \citep{mc05}. As center coordinates we
adopted $\alpha_{{\rm J}2000}=00:42:44.3$ (hh:mm:ss.s), $\delta_{{\rm
J}2000}=+41:16:09$ ($^{\circ}:':''$) and we used a position angle of
$\theta=38^{\circ}$ \citep{kent}. We find that clusters at projected
distances beyond 30 kpc from the galactic center, which corresponds to
the disk boundary defined by \citet{rac91}, are composed of two
components, i.e., $\sim$1 and $>$10 Gyr-old populations. These
clusters are real halo clusters even when taking into account
projection effects. \citet{mac10} studied the M31 halo GCs and
concluded that the majority of the halo clusters beyond a projected
radius of 30 kpc were accreted from satellite galaxies. However, it is
more likely that the older, $>$10 Gyr-old clusters were created when
the galaxy formed, while the younger $\sim$1 Gyr-old clusters might
either have been captured from satellite galaxies or created by
mergers and interactions in the last few Gyr. (In principle, a
homogeneous data set of cluster metallicities could help distinguish
between these scenarios. Unfortunately, the quality of metallicity
determinations, either from the literature or based on our SED fits,
for our sample clusters is insufficient for this purpose.) On the
other hand, for the clusters with projected radii $< 30$ kpc (the disk
boundary), we cannot easily distinguish whether they are disk clusters
or halo objects projected onto the disk. The ages of the star clusters
projected onto the disk range from $\sim 10^6$ to $> 10^{10}$ yr.

\begin{figure*}
\centerline{
\includegraphics[scale=.5,angle=0]{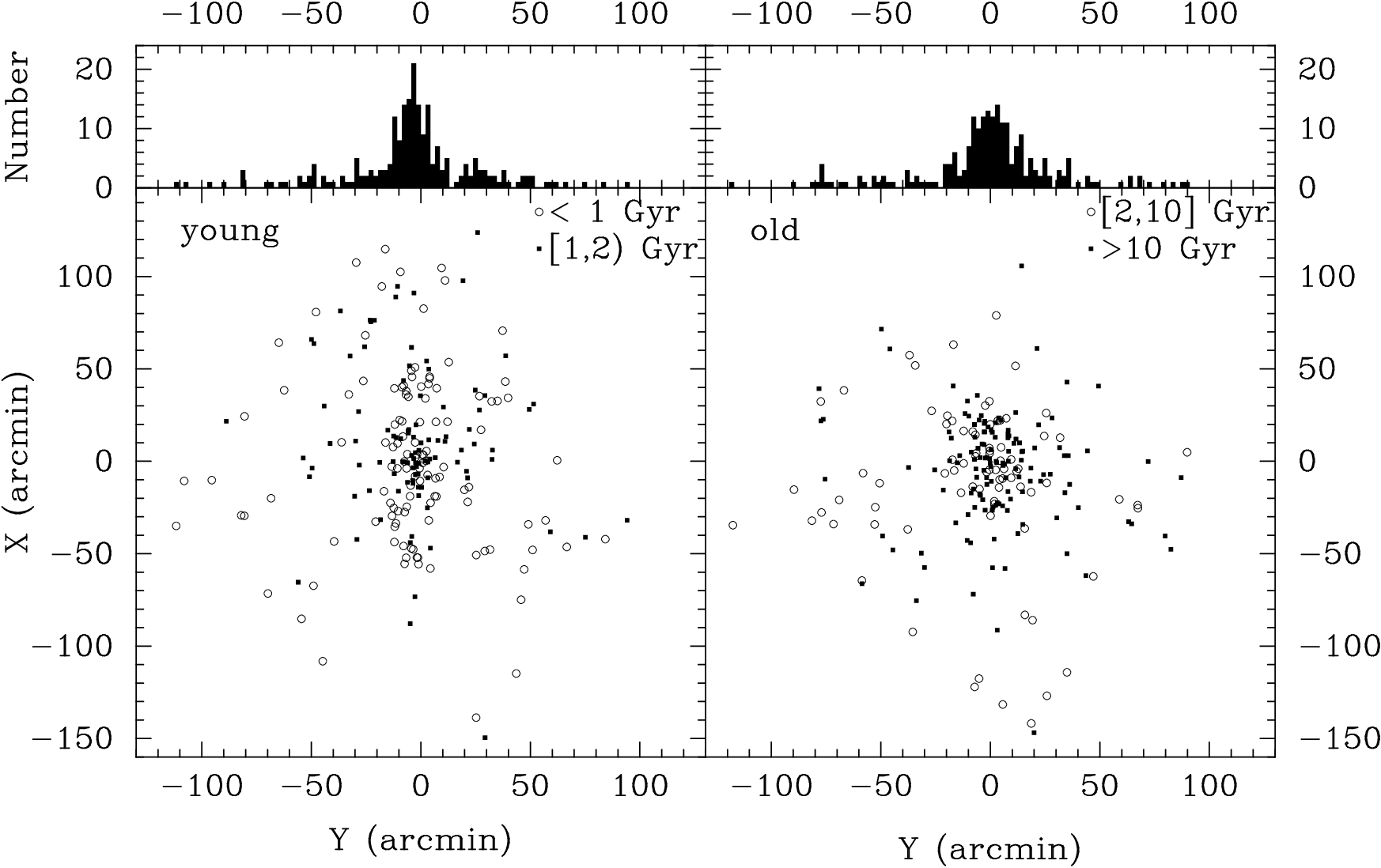}}
\caption[]{Galactocentric spatial distribution of (left-hand panel)
  the young (age $<$2 Gyr) and (right-hand panel) old (age $>$2 Gyr)
  globular-like star clusters in M31.}
\label{fig14}
\end{figure*}

Fig. \ref{fig14} shows the galactocentric spatial distribution of
the young, $<$2 Gyr-old (left-hand panel) and old, $>$2 Gyr-old star
clusters (right-hand panel). In the left-hand panel, the squares
represent the clusters younger than 1 Gyr, while in the right-hand
panel, the circles are clusters older than 10 Gyr. The $X$
coordinate is defined as the position along the major axis, while
the $Y$ coordinate represents the distance perpendicular to the
major axis \citep[see, e.g.,][]{perrett,wang10}. Formally,
\begin{equation}
X= A \sin\theta + B \cos\theta,
\end{equation}
\begin{equation}
Y=-A \cos\theta + B \sin\theta,
\end{equation}
\begin{equation}
A=\sin(\alpha-\alpha_0) \cos\delta,
\end{equation}
and
\begin{equation}
B=\sin\delta \cos\delta_0 -\cos(\alpha-\alpha_0) \cos\delta
\sin\delta_0.
\end{equation}

\begin{figure*}
\centerline{
\includegraphics[scale=.6,angle=0]{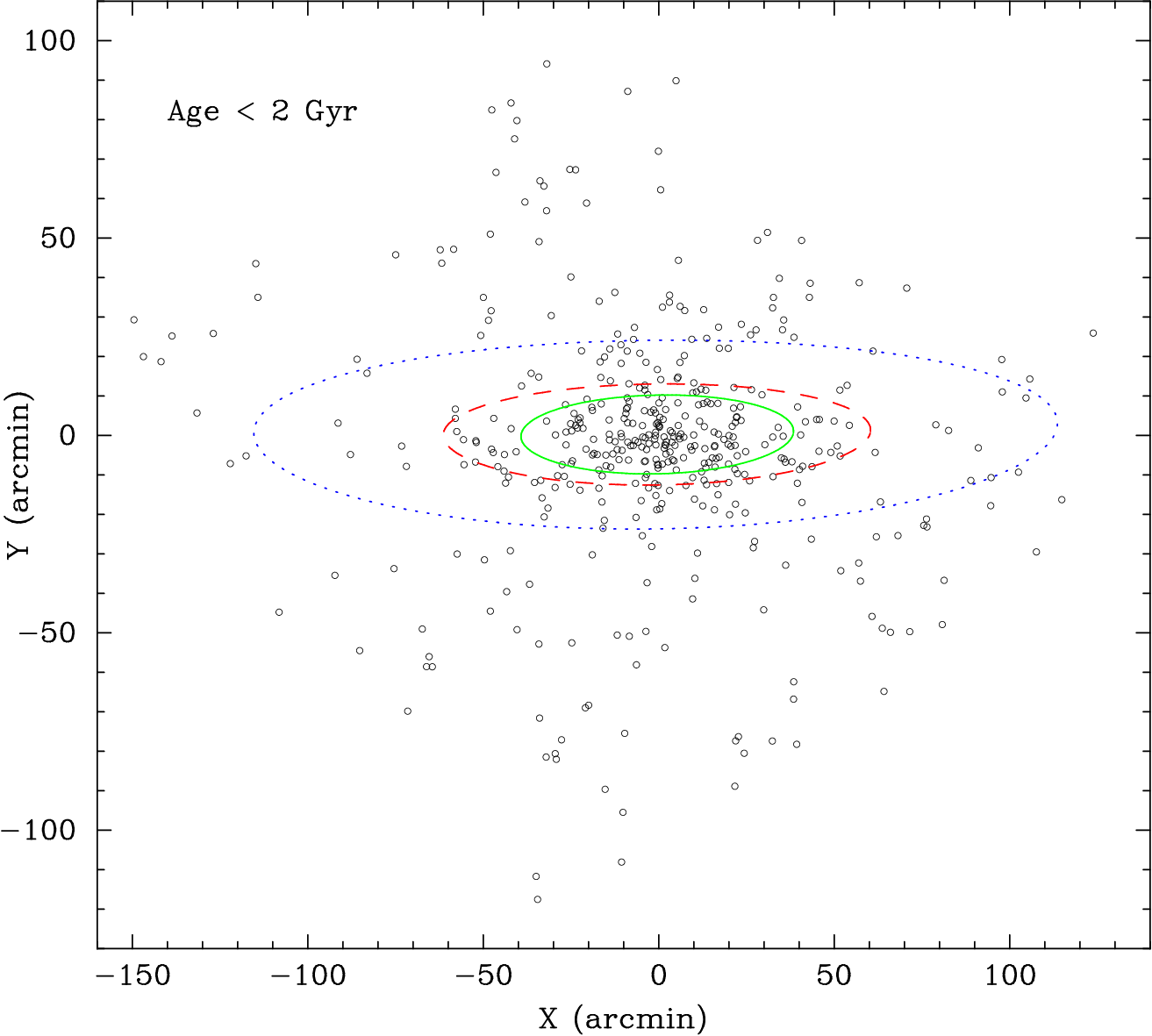}}
\caption[]{Galactocentric spatial distribution of the young (age $<$2 Gyr)
  globular-like star clusters in M31. There are 221 young clusters younger
  than 2 Gyr in our sample. The green ellipse (solid line) represents
  the 10 kpc ring, while the red ellipse (dashed line) represents the
  outer ring \citep{gord06}. The blue ellipse
  (dotted line) shows the disk boundary defined by \citet{rac91}.}
\label{fig15}
\end{figure*}

The red, solid ellipse and the black, dashed contour in the left-hand
panel of Fig. \ref{fig14} represent the `10 kpc ring' and the `outer
ring' of \citet{gord06} based on infrared observations with the {\sl
Spitzer Space Telescope}'s MIPS (Multiband Imaging Photometer for
Spitzer) instrument. In the 10-kpc and outer-ring regions, star
formation is very active, and hundreds of YMCs have been created over
the last few Gyr. The blue dotted ellipse represents the disk boundary
of M31 \citep{rac91}. We find that the YMCs in our sample are
spatially coincident with the disk and the spiral arms (for which we
use the 10 kpc and outer rings as proxy, respectively), and that the
majority of the young clusters (154 out of 221 clusters, 70\%) are
located in the disk defined by \citet{rac91}. Note that (in the
right-hand panel) the old clusters are located in the galaxy's bulge
and halo, which were both presumably created at the epoch when M31
formed. We also plot the histograms in the $y$ direction to show
the different distributions of the two subsamples. In the left-hand
panel, we find the highest-density region in the galaxy center,
corresponding to the bulge. There are two lower-density peaks nearby,
which may correlate with the 10 kpc `ring of fire' in
Fig.~\ref{fig15}. In the right-hand panel, the distribution is
completely different: the density is not as high as that of the young
population and there are no peaks nearby. This statistical test
implies that the spatial distributions of the two populations are
different and that the distribution of the young clusters correlates
with the galaxy's ring structure.

To show the spatial distribution of the young clusters and its
correlation with the ring structures more clearly, we include
Fig.~\ref{fig15}. The green ellipse (solid line) represents the 10 kpc
ring, while the red ellipse (dashed line) represents the outer ring
\citep{gord06}. The blue ellipse (dotted line) shows the disk boundary
defined by \citet{rac91}.

\citet{lam05a} show the characteristic cluster disruption time versus
ambient density in the disks of M33, the Small Magellanic Cloud, the
Milky Way, and M51. We also added the M82 data point from
\citet{deg05b}. The solid line is the theoretical prediction from
\citet{bm03} and the dashed line is from \citet{pz01}. It is interesting
to add M31 to this figure. For the disk of M31, we adopted a mass of
$7.0\times10^{10}M_{\odot}$ \citep{yin09} and an inner disk radius
given by the 10 kpc ring, while the outer disk boundary used is
$r_{\rm proj}=29.7$ kpc \citep{rac91}. The disk thickness adopted is
1.01 kpc \citep{ma97}. If we use the 10 kpc ring as the disk's size,
its mean density is $0.025 M_{\odot}$ pc$^{-3}$. However, if we use
$r_{\rm proj}=10$ kpc as disk size, the mean disk density is $0.221
M_{\odot}$ pc$^{-3}$, thus giving us a reasonable uncertainty
range. The characteristic cluster disruption time can be derived from
Eq. (8) of \citet{lam05a}. We obtained this disruption timescale, for a
cluster of mass $10^4 M_{\odot}$, $\log(t_4/{\rm yr})=9.47\pm0.24$
($2.95^{+2.18}_{-1.25}$ Gyr).

\begin{figure}
\centerline{
\includegraphics[scale=.35,angle=-90]{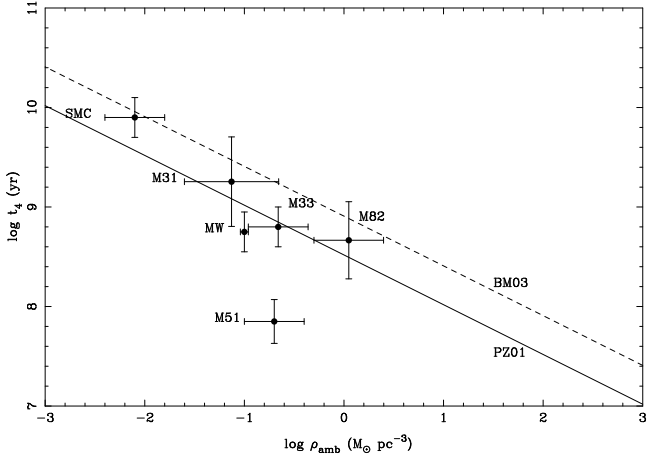}}
\caption[]{Characteristic cluster disruption time (for $10^4 M_\odot$
  clusters) versus ambient disk density for five local galaxies.  The
  solid and dashed lines are the theoretical predictions from
  \citet{bm03} and \citet{pz01}, respectively. For M33, the Small
  Magellanic Cloud (SMC), the Milky Way (MW), and M51, the data is
  from \citet{lam05a}. For M82, the data is from \citet{deg05b}.}
  \label{fig16}
\end{figure}

\begin{figure}
\centerline{
\includegraphics[scale=.35,angle=0]{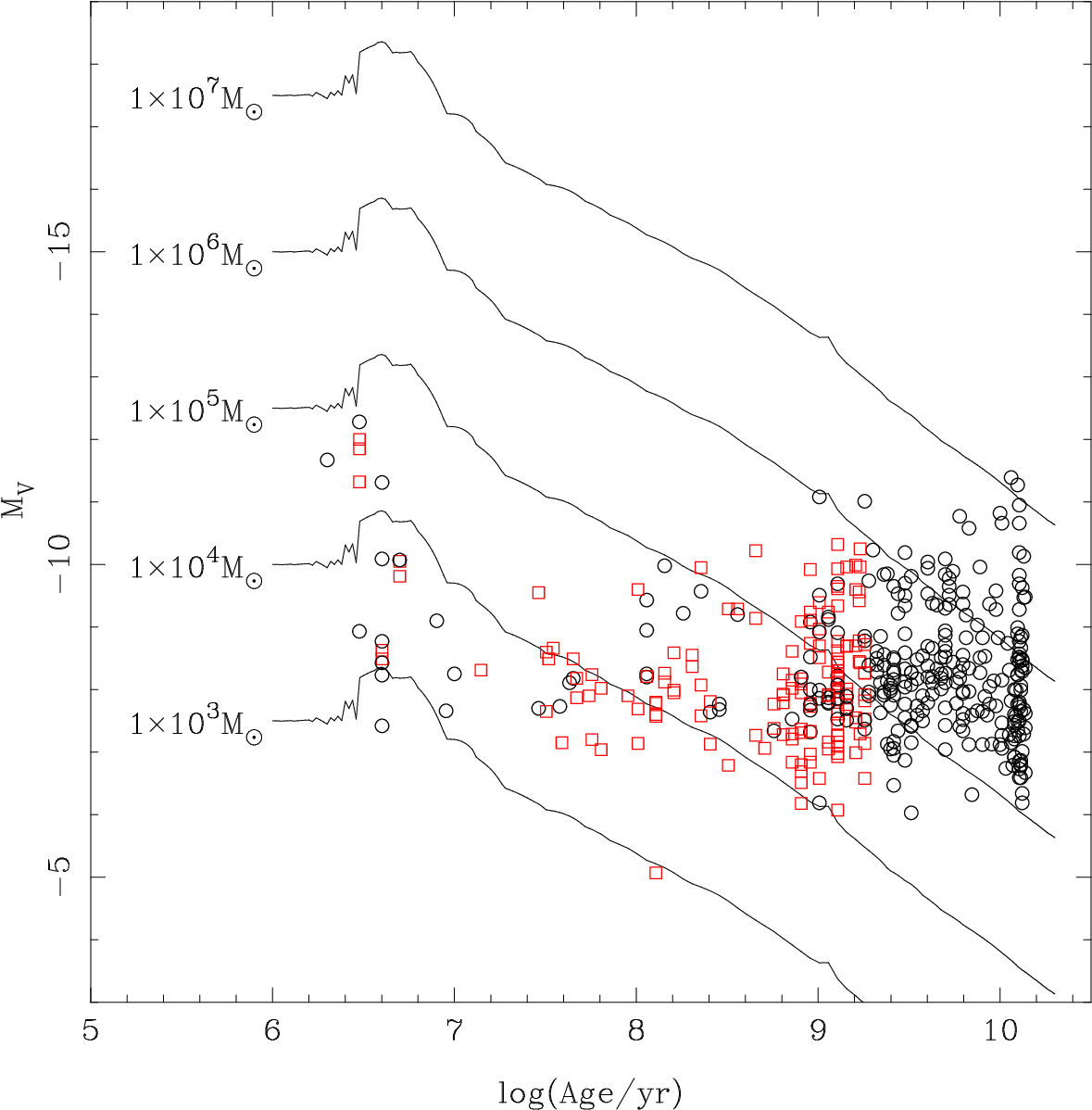}}
\caption[]{Dereddened, integrated $V$ magnitudes and initial cluster
  (SSP) masses as a function of age for our M31 sample clusters. The
  red squares represent the disk YMCs ($<2$ Gyr) identified in
  Fig.~\ref{fig10}, while the black circles are the other clusters,
  i.e., both the old clusters ($>2$ Gyr) and the young halo clusters.
  The continuous lines are fixed-stellar-mass models from \citet{bc03}
  for SSPs of solar metallicity, a \citet{chab} IMF, and Padova 1994
  evolutionary tracks. Disk YMCs of mass $10^4 M_\odot$ are expected
  to dissolve within $2.95^{+2.18}_{-1.25}$ Gyr and will not evolve to
  become GCs. However, for clusters located in the halo, the ambient
  density is much lower, so that they could survive for much longer.}
  \label{fig17}
\end{figure}

Given the large population of YMCs in M31, it is interesting to
assess how long they will likely survive and whether they may
eventually become old GCs. \citet{lam05a,lam05b} and \citet{lg06}
suggest that low-mass star clusters, in particular, will be
disrupted by stellar evolution, tidal stripping, spiral-arm
shocking, and encounters with giant molecular clouds, especially in
galactic disks. Fig. \ref{fig17} shows the dereddened, integrated
$V$-band magnitude versus age diagram for our sample clusters, with
theoretical predictions for SSP evolution for given initial masses
overlaid. The (red) open squares represent the disk YMCs identified
in Fig. \ref{fig15} and the continuous lines are fixed-stellar-mass
BC03 models for SSPs of solar metallicity, a \citet{chab} IMF, and
Padova 1994 evolutionary tracks. The black circles are the young
halo clusters and clusters older than 2 Gyr.

Using the disruption-time equation of \citet{lam05a}, we calculated the
relevant disruption times for the young disk clusters in our
sample. For clusters with a total mass $M_{\rm cl} = 10^3M_{\odot}$,
the disruption time $t_{\rm dis}=0.71^{+0.52}_{-0.30}$ Gyr, and for
$M_{\rm cl}= 10^4$ and $10^5 M_\odot$, the corresponding times are
$t_{\rm dis}=2.95^{+2.18}_{-1.25}$ and $12.3^{+9.1}_{-5.2}$ Gyr,
respectively. Very crudely, this suggests that if the initial cluster
mass $M_{\rm cl}> 10^5 M_{\odot}$, even star clusters located in the
dense M31 disk can survive and evolve into GCs over a Hubble time. We
note that most disk YMCs in our sample (red squares) will disrupt in
$< 2.24$ Gyr, while all are expected to have been disrupted by $t<
8.9$ Gyr. This suggests that none of the M31 disk YMCs will evolve
into $> 10$ Gyr-old GCs. The predicted fate of the M31 YMCs agrees
reasonably well the conclusions of \citet{cald09}, although they
claimed that most of the young clusters would be disrupted in
approximately 1 Gyr, except for some compact and massive objects.
These authors assumed that the density of M31's `ring of fire' is
similar to that of solar neighborhood, where the disruption time is on
the order of 1.3 Gyr \citep{lam05a}. Thus, they estimated most of the
M31 YMCs can only survive for approximately 1 Gyr. However, it is more
likely that the average density of the entire disk is much lower than
that of the `ring of fire,' so that the corresponding disruption time
should be much longer. \citet{bar09} investigated the nature of 23 M31
YMCs and concluded that most of their sample clusters cannot survive
for more than $\sim$5 Gyr. This is consistent with our result, $t_{\rm
dis}=2.95^{+2.18}_{-1.25}$ Gyr for a $10^4 M_\odot$ cluster (cf. their
fig. 6).

\subsection{Masses of the globular-like clusters}
\label{s:mass}

The mass-to-light ratio ($M/L$) values obtained from the spectroscopic
age estimates can be combined with our $V$-band photometry to derive
masses for all observed M31 GCs. Reddening values are, of course, also
needed for extinction corrections.

\begin{figure*}
\centerline{
\includegraphics[scale=.45,angle=-90]{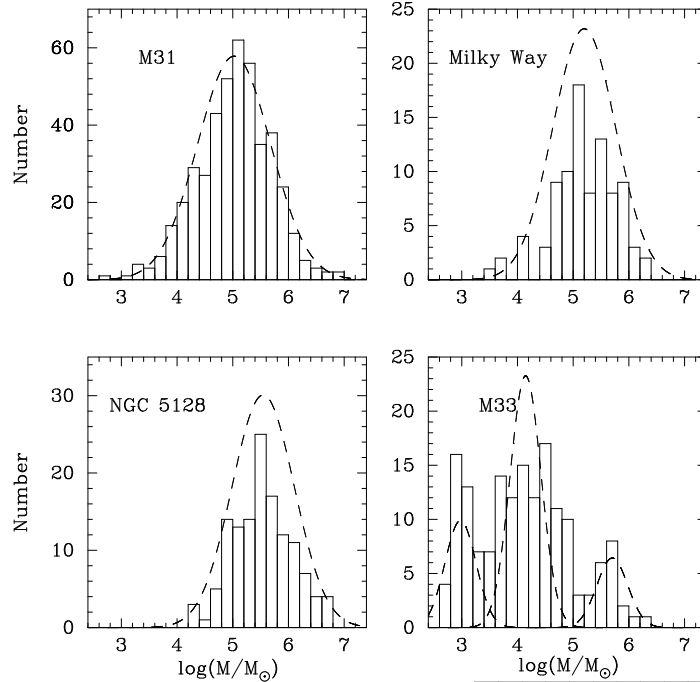}}
\caption[]{Mass distributions of globular-like clusters in M31 (top
  left-hand panel), and GCs in the Milky Way (top right-hand panel),
  NGC 5128 (bottom left-hand panel), and M33 (bottom right-hand
  panel).}
  \label{fig18}
\end{figure*}

We calculated the $M/L_V$ values using the BC03 models, luminosities
based on conversion of the $V$-band fluxes, and a distance modulus of
$(m-M)_0=24.47$ mag \citep{mc05}. The resulting masses are listed in
Table~\ref{agemass} as well as the metallicity and reddening
values applied in our fits. Fig. \ref{fig18} shows the mass
distribution for the confirmed M31 clusters in our sample. The peak
occurs at $\log(M_{\rm cl}/M_\odot)=5.03\pm0.02$, which is consistent
with the peaks in the mass distributions of the old GCs in M31. As can
be seen in Fig. \ref{fig18}, the lower-mass clusters ($M_{\rm cl} <
10^4 M_\odot$) might be either YMCs or OCs, which suggests that a
handful of clusters considered `GCs' in the RBC are, in fact, not old
massive M31 GCs. Fig. \ref{fig18} also shows the mass distributions of
the NGC 5128 clusters from \citet{mcl08}, the Milky Way GCs based on
King-model fits \citep{mv05}, and the M33 star clusters from
\citet{sara07}. The peak for the M31 cluster masses at $\log(M_{\rm
cl}/M_\odot)=5.03\pm0.02$ corresponds to $M_{\rm cl} \sim(1.07\pm
0.05)\times10^5 M_\odot$. For the Milky Way and NGC 5128 GCs, the
peaks occur at $\log(M_{\rm cl}/M_\odot)=5.20\pm0.03$ and
$5.55\pm0.03$, respectively, while for the M33 clusters the
distribution seems to be trimodal, with peaks at $\log(M_{\rm
cl}/M_\odot)=2.98, 4.15$, and 5.70, which is likely due to a mixture
of OCs, YMCs, and GCs in this sample.

\begin{figure}
\centerline{
\includegraphics[scale=.4,angle=-90]{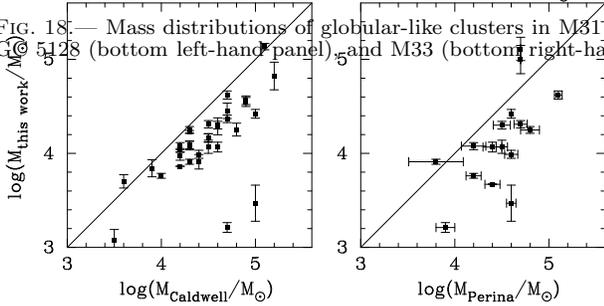}}
\caption[]{Comparisons of the clusters masses derived here with those
  from \citet{cald09} (left-hand panel) and \citet{per10} (right-hand
  panel).}
  \label{fig19}
\end{figure}

In Fig. \ref{fig19}, we compare our newly derived masses for the
confirmed, globular-like sample clusters with previous
determinations, where available (all for young clusters). The
left-hand panel is a comparison between \citet{cald09} and this
paper, while the right-hand panel shows a comparison between
\citet{per10} and our mass determinations. Note that the cluster
masses agree reasonably well among these three studies, although our
mass estimates are systematically smaller than those of both
\citet{cald09} and \citet{per10}. This systematic effect is most
likely due to the use of different methods \citep[cf.][]{deg05a}. In
this paper, we compared the continuum shapes (SEDs) with BC03 models
to derive the ages, and subsequently used the $M/L$s to obtain the
cluster masses. In contrast, \citet{cald09} only used the
spectral-line features and the Starburst99 SSP models \citep{sb99}
to constrain the cluster ages. \citet{per10} estimated cluster
masses based on integrated 2MASS $JHK$ photometry, combined with SSP
models. The latter authors found that different NIR magnitudes,
different IMFs, or different SSP models could all lead to different
mass determinations, with an overall accuracy of less than a factor
of three. Obviously, our results agree with \citet{per10} within
this uncertainly.

\begin{figure}
\centerline{
\includegraphics[scale=0.38,angle=-90]{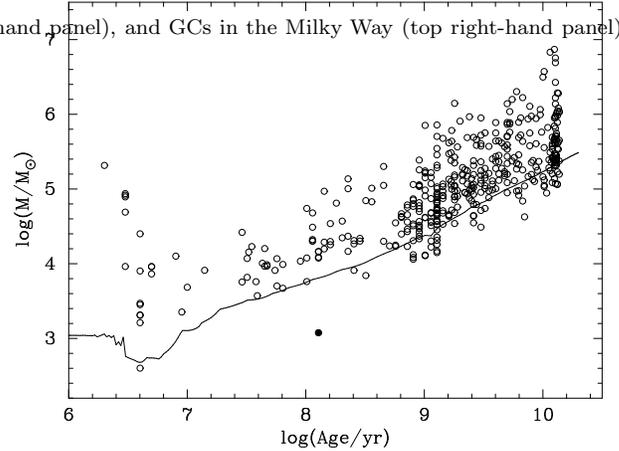}}
\caption[]{Mass versus age diagram for the confirmed M31 star clusters
  in our sample. The solid line is the fading (`completeness') limit
  based on a photometric sample cutoff at $M_V=-7$ mag (see
  Fig. \ref{fig6}) and BC03 SSP models of solar metallicity. The filled
  circle represents M078, which might be a young OC.}
  \label{fig20}
\end{figure}

Following \citet{da06} and \citet{wang10}, we investigated the
relationship between the ages and masses of the 445 confirmed
globular-like clusters and candidates in our sample (see Fig.
\ref{fig20}). The solid line is the `fading line' (roughly
equivalent to the $\sim 50$\% completeness limit) based on the BC03
SSP models for solar metallicity for an absolute magnitude of
$M_V=-7$ mag, derived from the half-peak point of Fig.~\ref{fig8}.
We note that most of the young clusters lie above the line. A number
of old clusters with ages between 1 and 10 Gyr are found below the
fading line, which is most likely due to their preferred loci in the
galaxy's halo, where fainter objects are more easily distinguishable
than when they are projected onto the bright disk. M078 (identified
by a filled circle) is the most extreme outlier. It is very faint
($V_0=19.40$ mag) and might, in fact, be a young OC. We also find a
number of YMCs \citep[as also noted by][]{cald09}, which might
survive to become old GCs by virtue of their high masses. Note that
there are two overdensity regions in this figure, at (i) $\log ({\rm
age/yr})\approx9.1$ ($\approx 1.28$ Gyr), temporally coincident with
the onset of thermally pulsing asymptotic giant branch stars around
an age of $\sim$ 1 Gyr; and (ii) at $\approx 13$ Gyr, which
represents a cluster population that seems to have formed during the
time of the galaxy's formation.

\section{Summary}
\label{s:sum}

We have presented an updated $UBVRI$ aperture-photometry catalog for
970 objects in the field of M31, including 965, 967, 965, 953, and 827
in the individual $UBVRI$ bands, respectively, of which 205, 123,
14, 126, and 109 do not have previously published photometry. These
objects, including globular-like clusters (GCs and YMCs), cluster
candidates, and other types of objects such as stars and galaxies,
were selected from the most comprehensive catalog of GCs and
candidates, the RBC v.4.0. We obtained our aperture photometry based
on archival images from the Local Group Galaxies Survey
\citep{massey}. We find good agreement between our photometry and
previous measurements, where available.

By combining our new $UBVRI$ photometry with the optical broad-band
$UBVRI$ and NIR $JHK$ magnitudes from the RBC v.4.0, we determined the
ages for 445 confirmed globular-like and candidate clusters in M31
using a $\chi^2$ minimization technique and employing BC03 theoretical
SSP synthesis models. Comparisons involving our sample clusters show
that our newly derived ages are consistent with previous
determinations. Over one half of these clusters are young, with ages
$<$ 2 Gyr, implying recent, active star formation in M31. This is
consistent with recent results invoking encounters and/or mergers with
other galaxies. This is quite different from the age distribution of
the Milky Way's GC system, implying different evolutionary histories
for M31 and the Milky Way.

The clusters in the halo ($r_{\rm proj}>30$ kpc) are composed of two
populations, old GCs with ages $>$ 10 Gyr and young, $\sim 1$ Gyr-old
clusters, suggesting that the globular-like clusters in M31 formed at
two different epochs. The spatial distribution of our sample clusters
suggests that YMCs with ages $<$ 2 Gyr are spatially coincident with
the disk (and a few with the halo) of M31 while the old star clusters
($>$2 Gyr) are spatially correlated with the bulge and halo. We find
that none of the young disk clusters can survive to become old GCs
after a Hubble time; instead, they will likely encounter giant
molecular clouds in the galaxy's disk and disperse as field stars on
timescales of a few Gyr.

We also estimated the masses of the 445 confirmed globular-like
clusters and candidates in our sample using the derived ages and BC03
theoretical $M/L$ ratios, combined with our new photometry from the
LGGS. The comparisons show that our estimates agree well with previous
results, where available. We calculated the characteristic disruption
timescales for the young disk clusters (age $<$ 2 Gyr) in M31 and
found that disk YMCs with a mass of $10^4 M_\odot$ are expected to
dissolve within 3.0 Gyr and will, thus, not evolve into old GCs.

\acknowledgments

This research was supported by the Chinese National Natural Science
Foundation through grants 10873016, 10803007, 10778720, 10633020,
10673012, 11003021 and 11043006, and by National Basic Research Program
of China (973 Program) under grant 2007CB815403. ZF acknowledges a Young
Researcher Grant of the National Astronomical Observatories, Chinese
Academy of Sciences.


\LongTables

\clearpage
\pagestyle{empty}
\begin{deluxetable}{ccccccc}
\tablecolumns{7} \tablewidth{0pc} \tablecaption{New $UBVRI$ photometry
  of M31 globular-like clusters.\label{photo}}

\tablehead{\colhead{Object} & \colhead{$U$}  &\colhead{$B$} &\colhead{$V$}  &\colhead{$R$}   &\colhead{$I$}  &\colhead{$r_{\rm ap}$} \\
\colhead{} &\colhead{mag}    &\colhead{mag} &\colhead{mag} &\colhead{mag} &\colhead{mag} &\colhead{\arcs}
}
\startdata
\input{photo.tex}
\enddata
\end{deluxetable}

\clearpage
\begin{deluxetable}{cccc}
\tablecolumns{4} \tablewidth{0pc} \tablecaption{Comparisons of our
  broad-band photometry with previously published M31 GC photometry
  from the RBC v.4.0. Offset = our measurement $-$
  literature value.\label{photocom}}
\tablehead{\colhead{Band} &\colhead{ Mean Offset } &\colhead{$\sigma$}
    &\colhead{Number} }
\startdata
\input{photocom.tex}
\enddata
\end{deluxetable}

\begin{deluxetable}{cccc}
\tablecolumns{4} \tablewidth{0pc} \tablecaption{Comparisons of our
  broad-band photometry with previously published M31 GC photometry
  from \citet{barmby}. Offset = our measurement $-$
  literature value.\label{photocomb}}
\tablehead{\colhead{Band} &\colhead{ Mean Offset } &\colhead{$\sigma$}
  &\colhead{Number} }
\startdata
\input{photocomb.tex}
\enddata
\end{deluxetable}

\begin{deluxetable}{cccc}
\tablecolumns{4} \tablewidth{0pc} \tablecaption{Comparisons of our
  broad-band photometry with previously published M31 GC photometry
  from \citet{peac10}. Offset = our measurement $-$
  literature value.\label{photocomp}}
\tablehead{\colhead{Band} &\colhead{ Mean Offset } &\colhead{$\sigma$}
  &\colhead{Number} }
\startdata
\input{photocomp.tex}
\enddata
\end{deluxetable}

\clearpage
\pagestyle{empty}
\begin{deluxetable}{ccccc}
\tablecolumns{5} \tablewidth{0pc} \tablecaption{Ages and masses of M31
  globular-like clusters.
\label{agemass}}
\tablehead{\colhead{Object} &\colhead{Age} & \colhead{${\rm
log(M_{cl})}$} &\colhead{[Fe/H]}&\colhead{E(B--V)}\\\colhead{}
&\colhead{Gyr} &\colhead{[${\rm M_{\odot}}$]} & \colhead{} &
\colhead{mag}} \startdata \input{result.tex}
\enddata
\tablenotetext{a} {Reddening obtained from free fits and not
available from \citet{fan08} or \citet{barmby}}.
\end{deluxetable}

\end{document}

%% file: photo.tex
 B001       & $ 18.783\pm  0.009$ &$ 18.261\pm  0.008$ &$ 17.156\pm  0.004$ &$ 16.455\pm  0.003$ &$ 15.712\pm  0.002$ &   2.90\\
 B005       & $ 16.880\pm  0.004$ &$ 16.552\pm  0.004$ &$ 15.529\pm  0.002$ &$ 14.895\pm  0.002$ &$   ...  ... $ &   5.13\\
 B015       & $ 20.344\pm  0.023$ &$ 19.411\pm  0.013$ &$ 18.130\pm  0.007$ &$ 17.265\pm  0.005$ &$ 16.381\pm  0.004$ &   2.19\\
 B017       & $ 17.547\pm  0.004$ &$ 17.048\pm  0.004$ &$ 15.968\pm  0.002$ &$ 15.246\pm  0.002$ &$   ...  ... $ &   3.86\\
 B018       & $ 18.614\pm  0.010$ &$ 18.372\pm  0.010$ &$ 17.517\pm  0.007$ &$ 16.961\pm  0.006$ &$ 16.351\pm  0.007$ &   3.86\\
 B021       & $ 19.301\pm  0.015$ &$ 18.764\pm  0.012$ &$ 17.650\pm  0.009$ &$ 16.904\pm  0.007$ &$ 16.048\pm  0.005$ &   3.86\\
 B025       & $ 17.959\pm  0.005$ &$ 17.702\pm  0.006$ &$ 16.799\pm  0.004$ &$ 16.205\pm  0.004$ &$   ...  ... $ &   2.90\\
 B027       & $ 16.669\pm  0.003$ &$ 16.486\pm  0.003$ &$ 15.642\pm  0.002$ &$ 15.104\pm  0.002$ &$   ...  ... $ &   3.86\\

%% file: photocom.tex
  U& $ -0.024\pm 0.016$ &  0.398 & 586 \\
  B& $  0.088\pm 0.015$ &  0.373 & 651 \\
  V& $  0.058\pm 0.011$ &  0.346 & 946 \\ 
  R& $  0.083\pm 0.014$ &  0.353 & 606 \\ 
  I& $  0.049\pm 0.013$ &  0.325 & 590 \\ 

%% file: photocomb.tex
  U& $  0.012\pm 0.018$ &  0.196 & 124 \\
  B& $  0.010\pm 0.015$ &  0.201 & 184 \\
  V& $  0.013\pm 0.014$ &  0.194 & 198 \\ 
  R& $  0.018\pm 0.013$ &  0.177 & 174 \\ 
  I& $  0.088\pm 0.022$ &  0.262 & 139 \\

%% file: photocomp.tex
  U& $ -0.014\pm 0.013$ &  0.305 & 594 \\
  B& $ -0.006\pm 0.008$ &  0.231 & 752 \\
  V& $  0.058\pm 0.007$ &  0.182 & 751 \\ 
  R& $  0.129\pm 0.009$ &  0.250 & 738 \\ 
  I& $  0.118\pm 0.014$ &  0.357 & 660 \\

%% file: result.tex
AU010     & $  8.000\pm  1.000$ & $  5.195\pm  0.059$ & $  0.077\pm  0.057$ & $  0.040\pm  0.010\tablenotemark{a}$ \\
B001      & $  5.750\pm  0.875$ & $  5.421\pm  0.062$ & $ -0.178\pm  0.085$ & $  0.250\pm  0.020$ \\
B002      & $  0.045\pm  0.017$ & $  4.203\pm  0.084$ & $  0.531\pm  0.099$ & $  0.500\pm  0.040\tablenotemark{a}$ \\
B003      & $  6.250\pm  2.000$ & $    nan\pm    nan$ & $ -2.249\pm  0.156$ & $  0.190\pm  0.020$ \\
B004      & $ 12.500\pm  5.500$ & $  5.469\pm  0.159$ & $ -0.490\pm  0.085$ & $  0.070\pm  0.020$ \\
B005      & $  2.300\pm  0.300$ & $  5.688\pm  0.052$ & $ -0.632\pm  0.099$ & $  0.280\pm  0.020$ \\
B006      & $  9.500\pm  2.625$ & $  5.995\pm  0.114$ & $ -0.405\pm  0.085$ & $  0.090\pm  0.020$ \\
B008      & $  0.905\pm  0.148$ & $  5.387\pm  0.039$ & $ -2.022\pm  0.156$ & $  0.640\pm  0.040\tablenotemark{a}$ \\